\documentclass[revtex4,apj]{emulateapj-rtx4}

\usepackage{lscape}

\newcommand{\eg}{e.g., }
\newcommand{\ie}{i.e., }
\newcommand{\Msun}{{\rm M}_{\odot}}

\newcommand{\Nifs}{$^{56}$Ni}

\newcommand{\Mms}{M_{\rm MS}}
\newcommand{\Mpeak}{M^{\rm bol}_{\rm peak}}
\newcommand{\fej}{f_{\rm ej}}

\newcommand{\E}{E_{\rm 51}}
\def\gsim{\mathrel{\rlap{\lower 4pt \hbox{\hskip 1pt $\sim$}}\raise 1pt
\hbox {$>$}}}
\def\lsim{\mathrel{\rlap{\lower 4pt \hbox{\hskip 1pt $\sim$}}\raise 1pt
\hbox {$<$}}}
\newcommand{\MCini}{M_{\rm cut}}

\newcommand{\Mmout}{M_{\rm mix}}
\newcommand{\Mni}{M{\rm (^{56}Ni)}}

\newcommand{\ye}{Y_{\rm e}}

\newcommand{\Mbh}{M_{\rm rem}}

\newcommand{\frho}{f_{\rm \rho}}

\newcommand{\Mfe}{M_{\rm ej}{\rm (Fe)}}
\newcommand{\Mec}{M_{\rm ej}{\rm (C+N)}}
\newcommand{\Mc}{M_{\rm pre}{\rm (C+N)}}

\newcommand{\Mrem}{M_{\rm rem}}

\newcommand{\Mco}{M_{\rm C+O}}

\slugcomment{Accepted for publication in the Astrophysical Journal.}

\begin{document}

\title{Abundance Profiling of Extremely Metal-Poor Stars and Supernova
Properties in the Early Universe}

\author{
 Nozomu~Tominaga\altaffilmark{1,2},
 Nobuyuki~Iwamoto\altaffilmark{3},
 Ken'ichi~Nomoto\altaffilmark{2,4}
 }

\altaffiltext{1}{Department of Physics, Faculty of Science and
Engineering, Konan University, 8-9-1 Okamoto,
Kobe, Hyogo 658-8501, Japan; tominaga@konan-u.ac.jp}
\altaffiltext{2}{Kavli Institute for the Physics and Mathematics of the
Universe (WPI), The University of Tokyo, 5-1-5 Kashiwanoha, Kashiwa, Chiba
277-8583, Japan}
\altaffiltext{3}{Nuclear Data Center, Japan Atomic Energy Agency, Tokai,
Ibaraki 319-1195, Japan; iwamoto.nobuyuki@jaea.go.jp}
\altaffiltext{4}{Department of Astronomy, School of Science,
The University of Tokyo, Bunkyo-ku, Tokyo 113-0033, Japan;
nomoto@astron.s.u-tokyo.ac.jp}

\setcounter{footnote}{4}

\begin{abstract}
 After the Big Bang nucleosynthesis, the first heavy element enrichment
 in the Universe was made by a supernova (SN)
 explosion of a population (Pop) III star (Pop III SN). The
 abundance ratios of elements produced from Pop III SNe are recorded
 in abundance patterns of extremely metal-poor (EMP) stars. The
 observations of the increasing number of EMP stars have made it
 possible to statistically constrain the explosion properties of Pop III SNe. 
 We present Pop III SN models whose nucleosynthesis yields well-reproduce
 individually the abundance patterns of 48
 such metal-poor stars as [Fe/H]~$\lsim-3.5$.
 We then derive relations between the abundance ratios of EMP
 stars and certain explosion properties of Pop III SNe:
 the higher [(C+N)/Fe] and [(C+N)/Mg] ratios correspond to the smaller ejected Fe mass
 and the larger compact remnant mass, respectively. Using these relations,
 the distributions of the abundance ratios of EMP stars are
 converted to those of the explosion properties of Pop III SNe. Such
 distributions are compared with those of the explosion properties of
 present day SNe: The distribution of the ejected Fe mass of Pop III
 SNe has the same peak as that of the present day SNe but shows an extended
 tail down to $\sim10^{-2}-10^{-5}\Msun$, and the distribution of
 the mass of the compact remnant of Pop III SNe is as wide as that of the
 present day stellar-mass black holes. Our results demonstrate the
 importance of large
 samples of EMP stars obtained by ongoing and future EMP star surveys and
 subsequent high-dispersion spectroscopic observations in 
 clarifying the nature of Pop III SNe in the early
 Universe.
\end{abstract}


\section{INTRODUCTION}
\label{sec:intro}

Just after the Big Bang, the Universe was composed of only H, He, and a small amount of light
elements and its further chemical enrichment was provided by 
stars and supernovae (SNe). The early history of the chemical enrichment
is recorded in chemical abundance of long-lived low-mass stars that were formed
in the early Universe. Such old stars show low heavy element (metal)
abundances, thus being called metal-poor (MP) stars. 
A large number of candidates of
MP stars have been discovered by several large survey programs
(\eg HK survey: \citealt{bee92,bee99}; Hamburg/ESO survey:
\citealt{chr03}; SEGUE survey: \citealt{yan09}). 
High-dispersion spectroscopic observations of these candidates
have been made to accurately determine their metallicities and detailed
elemental abundances (e.g., \citealt{rya96}, see also \citealt{yon13}
and the series for recent reviews). 

The first enrichment of heavy elements in the Universe was provided by supernova
(SN) explosions of population
(Pop) III stars (Pop III SNe), which ejected synthesized metals.
The next-generation stars formed from gases of mixture of the SN ejecta and
pristine interstellar matter. Metal abundance ratios in those
second-generation stars directly
reflect nucleosynthesis in the Pop III SNe. The ejecta of subsequent SNe
overlapped with each other, from which later-generation stars
formed. After many cycles of such star formation and SN explosions, the
heavy elements from numerous SN ejecta were well-mixed in the
interstellar gas and, eventually, the interstellar medium (ISM) became
chemically homogeneous. In the Galactic halo, such a transition from the
chemically inhomogeneous to homogeneous ISM took place at 
[Fe/H]\footnote{Here [A/B] 
$= \log_{10}(N_{\rm A}/N_{\rm B})-\log_{10} (N_{\rm A}/N_{\rm B})_\odot$, 
where the subscript $\odot$ refers to the solar value and $N_{\rm A}$
and $N_{\rm B}$ are abundances of elements A and B,
respectively.}~$\sim-3$ to $-2.5$ as suggested from the changes in the abundance
patterns of the MP stars, especially $r$-process elements, and the
hierarchical chemical evolution models (\eg \citealt{arg00,tum06}). 

In the inhomogeneous environment, chemical enrichment by a single SN dominates
the preexisting metal contents (\eg \citealt{aud95}) and [Fe/H] of an MP
star is not a
time indicator but merely reflects [Fe/H] of a cloud in which the MP
star formed. A single SN enriches the ISM roughly
up to [Fe/H]~$\sim-3$ \citep{tom07b,nom13}. Thus, MP
stars with [Fe/H]~$<-3$, called extremely metal-poor (EMP) stars
\citep{bee05}, are regarded to possess
nucleosynthetic products of a few Pop III SNe. 
Inversely, comparisons between nucleosynthetic yields of Pop III SNe and 
the abundance patterns of the EMP stars constrain explosion properties
of Pop III SNe (``{\sl abundance profiling}''). 

Asphericity and the explosion energy of Pop III SNe are particularly
important indicators obtained from such comparisons. 
To consistently reproduce [$\alpha$/Fe] and [(Co,~Zn)/Fe] of EMP
stars, materials synthesized in the deep complete Si-burning region
(including Co and Zn) should
be ejected but the amount of Fe should not be too large relative to $\alpha$
elements. This requirement is
satisfied with taking into account aspherical effects such as the
jet-like explosion \citep{tom07a,tom09a} or the Rayleigh-Taylor
instability \citep{jog09,jog10}. The aspherical explosion is
mimicked by the mixing-and-fallback model that well-reproduces the
abundance patterns of the EMP stars
\citep{ume02a,ume03,ume05a,iwa05,tom07b,heg10}. 
Furthermore, the trends of increasing [(Co,~Zn)/Fe] toward
lower [Fe/H] is shown to be explained if less-massive stars (with
main-sequence mass $\Mms\lsim20\Msun$) produce normal SNe and
more-massive stars (with $\Mms\gsim20\Msun$) produce energetic
explosions, \ie hypernovae \citep{ume05a,tom07b}.

The asphericity and the dependence of explosion energy ($E$) on $\Mms$
are also seen in the nearby present day SNe. Their nebular
spectra and polarization indicate aspherical distributions of elements
\citep[\eg][]{mae08,tanaka12}. And $\Mms$ and $E$ of the present
day SNe are constrained from their light curves
and spectra \citep[\eg][]{nom06}. The above variations of the present day
SNe are consistent with the requirements for the variations of Pop III
SNe to reproduce the abundance patterns and the trends of EMP stars. 
Such a consistency suggests that the SN explosion mechanism may not be
so sensitive to the metallicity. 

However, Pop III SN models for individual EMP stars have been so far constructed only for 
a few EMP stars which have interesting characteristics (\eg
\citealt{ume03,iwa05}). Also, because of the lack of large samples of
EMP stars, it has not been possible to investigate the
statistical properties of Pop III SNe.
Recently the number of the EMP stars increased dramatically as a result
of the large MP
star surveys, intensive follow-up high-dispersion spectroscopic
observations, and the development of stellar abundance analysis methods.
Therefore, it has become feasible and necessary to statistically
constrain the explosion properties of Pop III SNe using the EMP stars.

In this paper, we focus on the 48 most metal-poor stars\footnote{The stars
are found in ADS (\url{http://adswww.harvard.edu/}), SAGA database
(\url{http://saga.sci.hokudai.ac.jp/}, \citealt{sud08,sud11}), and
a comprehensive study on the EMP stars \citep{yon13}. We except the 
Si-deficient star, HE~1424-0241 \citep{coh07}, because its abundance
pattern can be reproduced only with an aspherical 
abundance distribution of the jet-like
explosion \citep{tom09a}.} with [Fe/H]~$\lsim-3.5$ including 41 halo
stars and 7 stars in dwarf galaxies, and construct Pop III
SN models to reproduce their abundance patterns. With reasonably
successful models, the abundance ratios of these EMP
stars can be directly related to the explosion properties of Pop III
SNe: the masses of ejected Fe $\Mfe$ and the compact remnant
$\Mrem$. According to these relations, we derive the distributions
of $\Mfe$ and $\Mrem$ of Pop III SNe. These relations make it possible to extract
information of Pop III SNe without constructing SN models at the
era that large samples of EMP stars are available.

We describe the model and procedure in \S~\ref{sec:model} and present SN
models in \S~\ref{sec:obs}. After discussing the limitations of our
approach in
\S~\ref{sec:limit}, the distributions of properties of Pop III SNe are
presented in \S~\ref{sec:dist}. The paper is concluded with several
discussion in \S~\ref{sec:discussion} and remarks in
\S~\ref{sec:conclusion}.

\begin{figure*}
\epsscale{1.}
\plotone{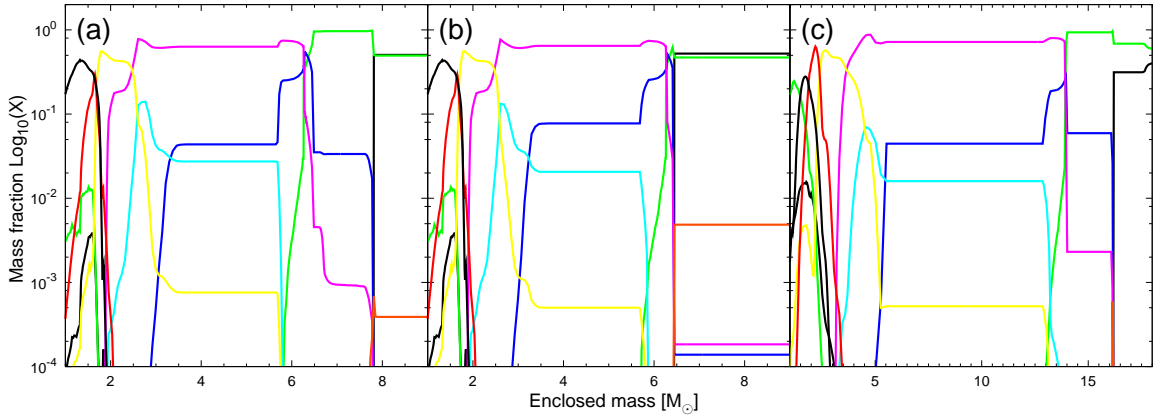}
\figcaption{Abundance distribution against the enclosed mass $M_r$
 before the explosion of (a) $25\Msun$ star without mixing
 enhancement (Model A), (b) $25\Msun$ star with mixing
 enhancement (Model B), and (c) $40\Msun$ star without mixing
 enhancement (Model C). The colors represent $^1$H ({\it black}), $^4$He
 ({\it green}), $^{12}$C ({\it blue}), $^{14}$N ({\it orange}), $^{16}$O
 ({\it magenta}), $^{24}$Mg ({\it cyan}), $^{28}$Si ({\it yellow}),
 $^{54}$Fe ({\it gray}), and $^{56}$Ni ({\it red}). \label{fig:preSN}}
\end{figure*}

\section{Model \& Method}
\label{sec:model}

We adopt the
$25\Msun$ progenitor models to
construct Pop III SN models for all EMP stars in a unified manner, and
the $40\Msun$ progenitor model to investigate dependence on $\Mms$.
The $25\Msun$ progenitor models with a different mixing efficiency are
taken from \cite{iwa05} and the $40\Msun$ progenitor model is taken from
\cite{tom07a} as follows:

\noindent (1) Model A is the $25\Msun$ progenitor without enhanced mixing,

\noindent (2) Model B is the $25\Msun$ progenitor with enhanced mixing, \eg due to rapid
rotation,
 
\noindent
and

\noindent (3) Model C is the $40\Msun$ progenitor without enhanced
mixing. 

The difference due to
the enhanced mixing prominently appears in the abundance of N that is synthesized
via the CNO cycle facilitated by the mixing of C in the He convective
shell into the H-rich envelope. Thus Models A and C have a low N
abundance, while Model B has a high N abundance in the H layer
(Figs.~\ref{fig:preSN}a-\ref{fig:preSN}c). The efficient mixing also considerably
enhances the He abundance in the H layer. The progenitor models are
summarized in Table~\ref{tab:preSN}. We adopt explosion energies 
$\E=E/(10^{51}~{\rm erg})=0.72,~1,~3,~5,~10,~{\rm and}~20$ for Models A
and B, and $\E=10,~15,~20,~{\rm and}~30$ for Model C. 

Hydrodynamical and nucleosynthesis calculations are performed with the
same method as in \cite{tom07b}. The explosion is initiated by a thermal
bomb and followed with the Lagrangian code assuming spherical symmetry.
The hydrodynamical calculation includes nuclear energy releases from the
$\alpha$-network and a realistic equation of state \citep{nom82,nom88}.
The detailed nucleosynthesis calculation is performed as a post-processing
\citep{hix96,hix99} with a reaction network including 280 isotopes up
to $^{79}$Br (Table~1 in \citealt{ume05a}). 

In order to take into account the aspherical effect, we apply the
mixing-and-fallback model to the spherical calculations. The
mixing-and-fallback model has three parameters characterizing the asphericity:
the initial mass cut $\MCini$, the outer boundary of the mixing region
$\Mmout$, and the ejection factor $\fej$ (Appendix in 
\citealt{tom07b}). The mass of the compact remnant $\Mrem$ is derived as 
\begin{equation}
  \Mrem=\MCini+\left(1-\fej\right)\left(\Mmout - \MCini\right). \label{eq:mfmodel}
\end{equation}
In the jet-induced explosion model, physical meanings of $\MCini$,
$\fej$, and $\Mrem$ are the inner boundary, the fraction of the solid angle of
a region from which materials produced by incomplete Si burning are
ejected, and the remnant mass, respectively \citep{tom09a}.
An explosion with weak jets, \ie with low energy deposition rate, corresponds
to an SN model with small $\fej$ and large $\Mrem$ \citep[][]{tom07a}. 

\cite{tom09a} showed that the mixing-and-fallback model do not
fully reproduce the enhanced abundance ratios [(Sc, Ti, Co,
Zn)/Fe] obtained in the jet-induced explosion model. The enhancement stems from
nucleosynthesis in a high-entropy environment which is realized by the
concentration of energy in the narrow jet. Thus, we
adopt ``low-density'' modification as in \cite{tom07b} to mimic the
high-entropy environment, in which the density of the progenitor is
artificially reduced by a factor of ``low-density'' factor $\frho$
without changing the mass of each Lagrangian mass shell. Furthermore, since the neutrino
absorption during the explosion may largely change the $\ye$
distribution in the Si burning layers \citep[\eg][for recent
review]{jan12,kot12,bur13,bru13}, which is not included in our explosion
calculations, we modify $\ye$ profile to $Y_{\rm e}=0.5001$ in the
complete Si burning region if [Co/Fe] of an SN model is underabundant and/or 
$Y_{\rm e}=0.4997$ in the incomplete Si burning region if [Mn/Fe] of an
SN model is
underabundant \citep{ume05a}. Because the explosion models with
realistic geometry and the neutrino processes have not been established,
these modifications to compare with the observed abundance ratios of
many EMP stars may provide useful constraints on the entropy and
$\ye$ distribution in the realistic explosion models. We note that these
two modifications do not affect on the distribution of
$\Mfe$ and $\Mrem$ of Pop III SNe (\S~\ref{sec:dist}).

\section{Comparison with the observations}
\label{sec:obs}

Hydrodynamical and thermodynamical features of SNe appear in
abundance ratios between elements synthesized in different layers and
between elements synthesized in the same layer, respectively. It is
important to determine abundances of elements as many as possible because
different abundance ratios provide different information. We
focus on the hydrodynamical features in this paper and select the 48 EMP
stars (including a binary) which have at least 3
determined abundance ratios to Fe of the following elements: (1) C, N, or O,
(2) Mg, (3) Ca, and (4) Co or Ni. They consist of 19 carbon-normal and
nitrogen-normal EMP stars including a binary system, 12 carbon-enhanced
metal-poor (CEMP) stars, 6 nitrogen-enhanced metal-poor (NEMP)
stars, 2 ultra metal-poor (UMP) star, and 2 hyper metal-poor (HMP)
stars in Galactic halo and 7 EMP stars in dwarf galaxies. Definitions of
classification are described in respective subsections. The stars and
references are summarized in Table~\ref{tab:stars}. 

Since the theoretical model provides mass fractions of elements and
isotopes, the theoretical SN yields should be compared with genuine
abundance ratios taking into account non-LTE (NLTE)
and 3D effects. However, the NLTE or 3D effect has been estimated only for
several MP stars \citep[\eg][]{col06,and07,and08,and10} and 3D-NLTE
abundance ratios are currently not available for any stars. Results could be
systematically biased if we adopt 1D-LTE abundance ratios. On the other
hand, in order to investigate statistical nature of Pop III SNe, it
is requested to avoid including artificial scatter by adopting different stellar
atmosphere models, \ie to derive abundance ratios with a uniform
method. \cite{yon13} recently present homologous abundance analyses
and discuss distribution of abundance ratios but this study is
restricted within a framework of a 1D-LTE assumption. 

In this paper, we adopt abundance ratios in
\cite{yon13} and in other literatures which may take into
account NLTE or 3D effect (Table~\ref{tab:stars}). 
The abundance patterns of the EMP stars and the SN yields are
normalized with the solar abundance ratios in \cite{asp09}.
If there are plural references, we adopt their average
[Fe/H]. We construct an SN model to reproduce NLTE or 3D
corrected abundance ratios if the corrections specified for the individual EMP
star are available. If not, we construct an SN model to reproduce 1D-LTE
abundance ratios but also refer abundance ratios for which the following NLTE
correction is expediently applied: 
\begin{itemize}
 \item $\log(g)$ and [Fe/H]-dependent correction for [Na/Fe] (\citealt{and07},
see also \citealt{lin11}),
 \item $+0.2$ for [Mg/Fe] \citep{and10},
 \item $T_{\rm eff}$, $\log(g)$, and [Fe/H]-dependent correction for [Al/Fe]
\citep{and08},
 \item $-0.2$ for [K/Fe] \citep{tak09,and10},
 \item $+0.4$ for [Cr/Fe] derived from \ion{Cr}{1} line(s)
 \citep{ber10cr}, and 
 \item $+0.6$ for [Mn/Fe] derived from \ion{Mn}{1} resonance line(s)
\citep{ber08}. 
\end{itemize}
The corrections for [Na/Fe] and [Al/Fe] are estimated from the
interpolation and extrapolation of the corrections given by
in \cite{and07,and08}, and the average corrections are
uniformly adopted for other elements. The true correction
depends on the stellar parameters and should be estimated for the
individual star. Thus, the range between 1D-LTE and expediently-corrected
ratios is taken as a range of uncertainty. 

\begin{figure*}
\epsscale{.9}
\plotone{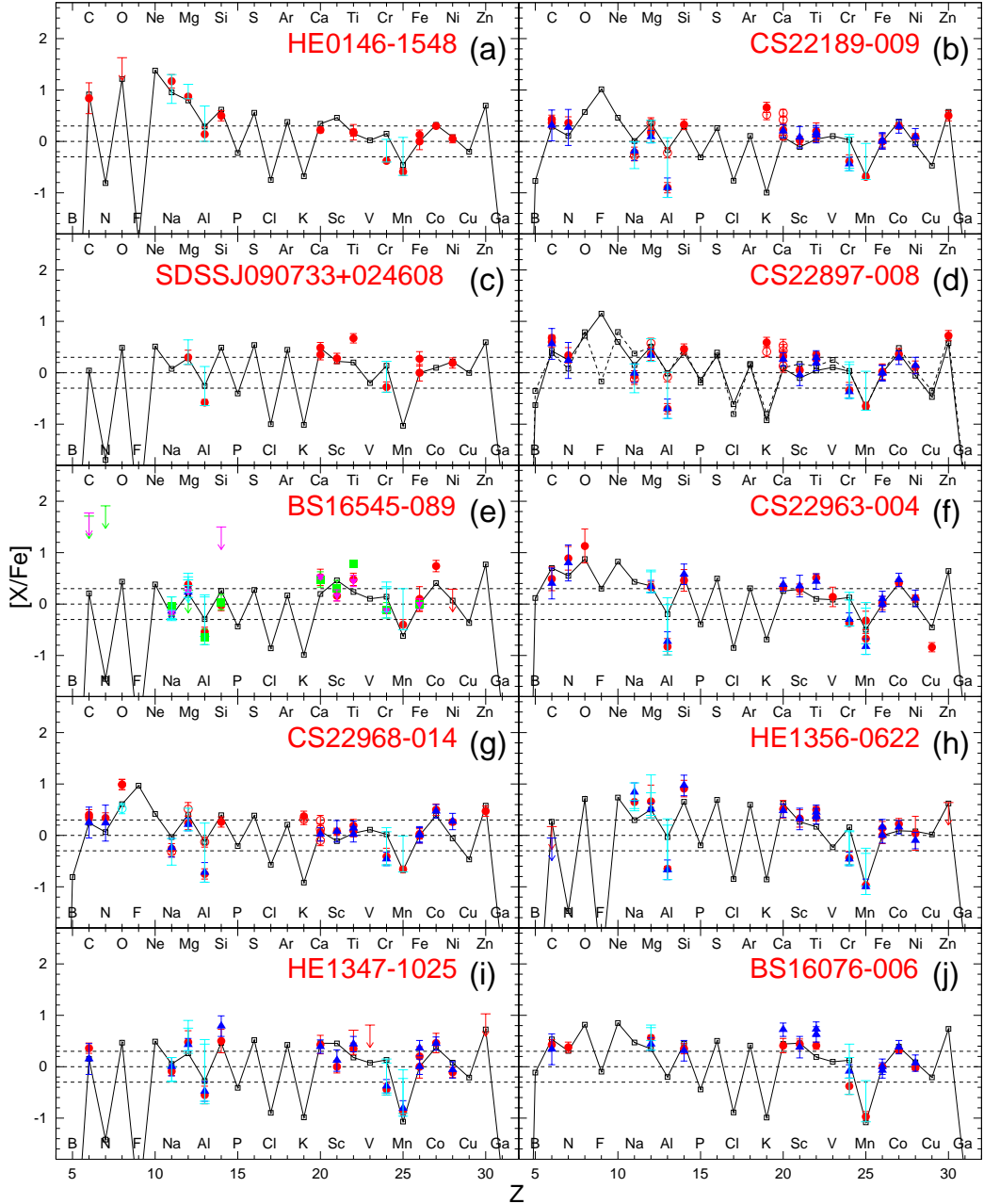}
\figcaption{Comparison between the abundance patterns of the individual EMP
 stars (1D-LTE: {\it red/blue/green/magenta/yellow filled points with bars}, 3D or NLTE:
 {\it red/blue/green/magenta/yellow open points with bars}, and
 expediently-corrected [see the text for the detail]: {\it cyan bars or
 open points with bars}) and the Pop III SN models ({\it lines with open
 squares}). The colors of points represent reference papers
 (see Table~\ref{tab:stars} in detail). The
 properties and legends of Pop III SN models are summarized
 in Table~\ref{tab:models}. \label{fig:EMP1}}
\end{figure*}

\begin{figure*}
\epsscale{.9}
\plotone{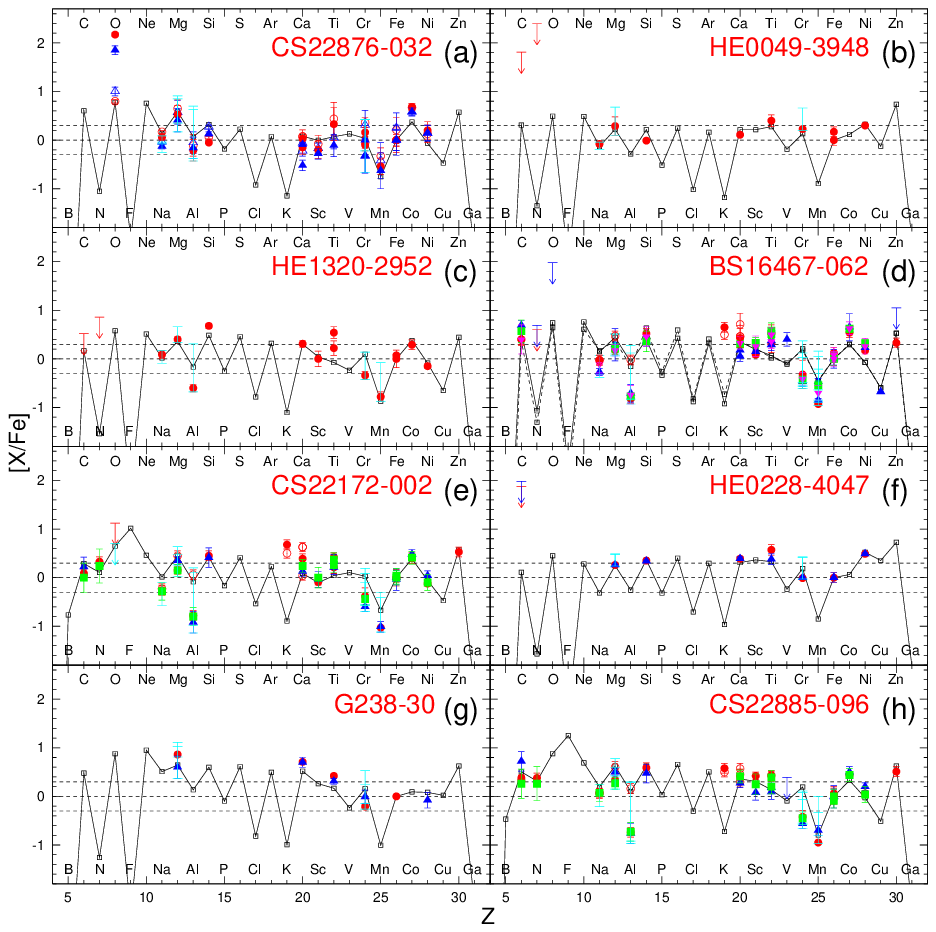}
\figcaption{Same as Fig.~\ref{fig:EMP1}. \label{fig:EMP2}}
\end{figure*}

\begin{figure*}
\epsscale{.9}
\plotone{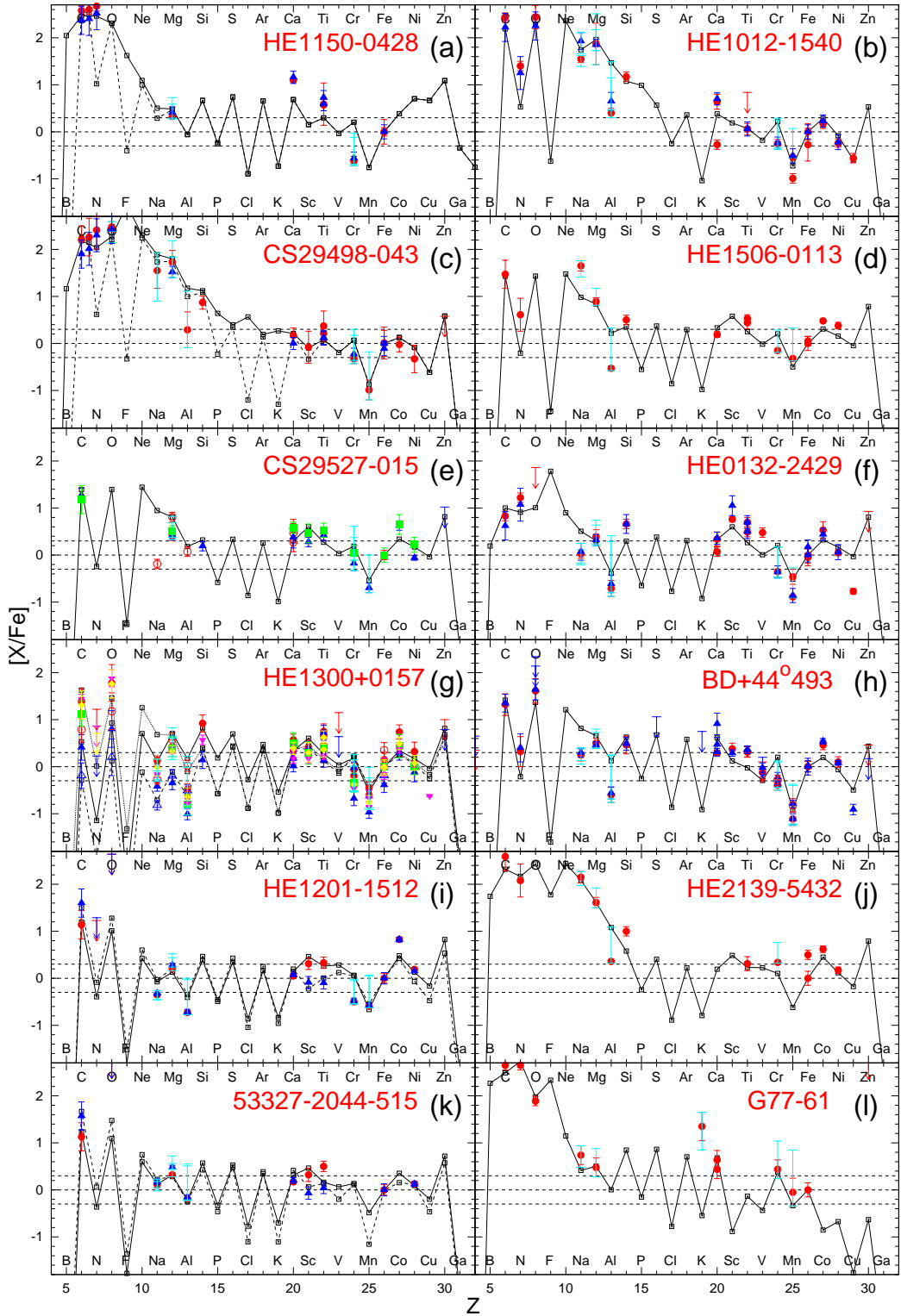}
\figcaption{Same as Fig.~\ref{fig:EMP1} but for the CEMP stars. \label{fig:CEMP}}
\end{figure*}

The properties of Pop III SN models are summarized in
Table~\ref{tab:models}. The comparisons between the abundance patterns
of the EMP stars and the SN yields are shown in
Figures~\ref{fig:EMP1}a-\ref{fig:EMP1}j and
\ref{fig:EMP2}a-\ref{fig:EMP2}h for the carbon-normal and nitrogen-normal EMP
stars, Figures~\ref{fig:CEMP}a-\ref{fig:CEMP}l for the CEMP stars,
Figures~\ref{fig:NEMP}a-\ref{fig:NEMP}f for the NEMP stars,
Figures~\ref{fig:UMP}a-\ref{fig:UMP}b for the UMP stars,
Figures~\ref{fig:HMP}a-\ref{fig:HMP}b for the HMP stars, and
Figures~\ref{fig:dSph}a-\ref{fig:dSph}g for the EMP stars in dwarf
galaxies.

\subsection{Extremely Metal-Poor stars}
\label{sec:EMP}

In this subsection, we examine 19 carbon-normal and nitrogen-normal EMP
stars with $-4\lsim$~[Fe/H]~$\lsim-3.5$. 

{\bf HE~0146--1548}: The metallicity is [Fe/H]~$=-3.46$ and the
abundance ratios are shown in Figure~\ref{fig:EMP1}a
\citep{yon13,nor13p4}. The abundance
pattern is well reproduced by an SN model with $\E=5$, $\Mbh=3.45\Msun$,
and $\Mfe=2.38\times10^{-2}\Msun$. 

{\bf $\star$ [Na/Mg]}: The
high [Na/Mg] ($>0$) is realized by the mixing region extending to the
middle of the O+Mg layer above which Na synthesized by static
burning is not destroyed by explosive burning. 

{\bf $\star$ [(Ti,~Co)/Fe]}: The high
[(Ti,~Co)/Fe] indicates that explosive nucleosynthesis takes place in
the high-entropy environment.

{\bf CS~22189--009}: The metallicity is [Fe/H]~$=-3.48$ and the
abundance ratios are shown in Figure~\ref{fig:EMP1}b
\citep{cay04,spi05,spi06,and07,and08,and10,spi12,yon13}. Since the star
shows [C/N]~$\sim0$ and the mixing in CS~22189--009 is not strong
\citep{spi05,spi06}, N is likely to be synthesized by the CNO
cycle in the SN
progenitor due to the enhanced mixing. The abundance pattern except for
[K/Fe] is well
reproduced by an SN model with $\E=20$, $\Mbh=4.20\Msun$,
and $\Mfe=7.76\times10^{-2}\Msun$. 

{\bf $\star$ [K/Fe]}: The underproduction of K could be solved
with nucleosynthesis in $p$-rich ejecta which may have been produced by neutrino
processes \citep[\eg][]{iwa06}. \cite{heg10} also presented that [K/Fe] could
be enhanced in some less-massive stars. 

{\bf $\star$ [F/Fe]}: The abundance of F is enhanced by the hot CNO cycle at the bottom of the
H layer during shock propagation. If an N-rich layer experiences $T\gsim7\times10^8$~K,
$^{19}$F is synthesized as
$^{14}$N($p$,$\gamma$)$^{15}$O($\alpha$,$\gamma$)$^{19}$Ne($\beta^+$)$^{19}$F.
Such an enhancement of F results if a progenitor star has 
the N-rich layer close to the core to experience high enough
temperature \citep[see also][]{chi12}. 
Thus [F/Fe] could give a clue to constrain N enhancement in the
progenitor\footnote{We note that F could not be enhanced in 
a red supergiant with the far-off N-rich layer even if N is enhanced in
the SN progenitor.} and the explosion
energy (\S~\ref{sec:energy}). 

{\bf $\star$ $^7$Li and $^{11}$B}: 
$^{11}$B is synthesized via
$^3$He($\alpha$,$\gamma$)$^7$Be($\alpha$,$\gamma$)$^{11}$C($\beta^+$)$^{11}$B
in the He-rich layer of the H envelope which experiences
$T\gsim5\times10^8$~K. The yield of $^7$Be decaying to $^7$Li is
$2\times10^{-7}\Msun$, which is smaller than the swept-up
mass of primordial $^7$Li synthesized by big bang
nucleosynthesis.\footnote{According to the swept-up H mass
(\citealt{tho98}, see also \citealt{tom07b}) and standard big bang
nucleosynthesis \citep[\eg][]{coc04}, the swept-up mass of primordial $^7$Li is
$\sim5.8\times10^{-6}\Msun$ for an SN explosion with $\E=1$.}

{\bf SDSS~J090733+024608}: The metallicity is [Fe/H]~$=-3.50$. The
abundance ratios are presented in \cite{caf11} but [(C,~N,~O)/Fe] are
not available (Fig.~\ref{fig:EMP1}c). The abundance
pattern except for [Ti/Fe] is well reproduced by an SN model with $\E=5$, $\Mbh=1.74\Msun$,
and $\Mfe=1.85\times10^{-1}\Msun$. It would be possible to reproduce the
abundance pattern without {\sl mixing-and-fallback} because of no
constraint on [C/Fe] and low [Mg/Fe]. On the other hand, the high
[(Sc,~Ti)/Fe] ratios indicate nucleosynthesis
in a high-entropy environment that is likely to be realized in an aspherical
explosion.

{\bf $\star$ [Ti/Fe]}: The disagreement in [Ti/Fe] could be solved by taking
into account the NLTE effect. The NLTE abundance determination by
\ion{Ti}{1} and \ion{Ti}{2} lines still needs to be improved.
However, if we adopt the lines of equal ionization stages as recommended
by \cite{ber11}, [\ion{Ti}{2}/\ion{Fe}{2}] ($=+0.4$) is 
consistent with the SN model.

{\bf CS~22897--008}: The metallicity is [Fe/H]~$=-3.50$ and the
abundance ratios are shown in Figure~\ref{fig:EMP1}d
\citep{cay04,spi05,spi06,and07,and08,and10,spi12,yon13}. 
The abundance pattern except for [K/Fe] is well reproduced by an
SN model with $\E=20$, $\Mbh=4.22\Msun$, and
$\Mfe=5.65\times10^{-2}\Msun$ (solid line).
The abundance pattern and thus the SN model are similar to those
of CS~22189--009. A model with lower explosion energy of $\E=5$ (dashed
line) also
reproduces the abundance pattern, except for the worse agreement of Na. 
The
distinct difference between the models with $\E=20$ and $5$ is seen in
the abundance of F because the maximum temperature of the N-rich layer
is higher for the higher explosion energy (\S~\ref{sec:energy}).

{\bf BS~16545--089}: The metallicity is [Fe/H]~$=-3.50$ and the
abundance ratios are shown in Figure~\ref{fig:EMP1}e 
\citep{coh04,lai08,aok09,yon13}. For CNO elements, only the loose upper
limits are obtained for [(C,~N)/Fe]. The [Ti/Fe] ratio given in
\cite{coh04} and \cite{lai08} is different, so its priority is low for
our model fitting. 
The abundance pattern except for [Co/Fe] is well reproduced by an SN model
with $\E=20$, $\Mbh=4.02\Msun$, and $\Mfe=1.07\times10^{-1}\Msun$. 

{\bf $\star$ [Co/Fe]}: [Co/Fe] is
enhanced by the low-density and $\ye$ modification but still
underabundant. It is known that [Co/Fe] has an NLTE effect \citep{ber10co}
but the correction is opposite to the improvement. Although the
[Co/Fe] can be enhanced as high as $\sim+0.9$ by nucleosynthesis in a
high-entropy and slightly $p$-rich hot bubble or accretion disk
\citep[\eg][]{pru04a,pru05}, [Co/Fe] in the integrated SN ejecta is
enhanced only up to $+0.4$ \citep[][]{fro06a,fro06b,izu10}. Such
high [Co/Fe] ($\sim+0.9$) is reproduced if the mixing of the SN ejecta is
inhomogeneous and a next-generation star forms from gases enriched only by a
portion of the SN ejecta with high [Co/Fe].

{\bf CS~22963--004}: The metallicity is [Fe/H]~$=-3.54$ and the
abundance ratios are shown in Figure~\ref{fig:EMP1}f
\citep{lai08,yon13}. The abundance pattern except for [(Ti,~Cu)/Fe] is
well reproduced by an SN model
with $\E=5$, $\Mbh=4.75\Msun$, and $\Mfe=2.85\times10^{-2}\Msun$. 

{\bf $\star$ [Cu/Fe]}: \cite{lai08} indicated the overproduction of Cu in the
SN model and proposed
that [Cu/Fe] in theoretical models can be reduced by appropriate choice of $\MCini$. However,
$\MCini$ is a fundamental parameter affecting other elements as
well. For example, the choice of $\MCini$ yielding lower [Cu/Fe] leads lower
[Zn/Fe]. On the other hand, the 3D effect of the stellar atmosphere
model reduces the observed [Cu/Fe] for a dwarf star
\citep{bon09cu}. This worsens the agreement. However, the effect for a
subgiant star as CS~22963--004 is not clarified and the effect is weaker
for lower metallicity. Furthermore, there has been no examination of the
NLTE effect on Cu. A 3D-NLTE study is needed to minutely investigate on Cu. 

{\bf $\star$ [O/Fe]}: [O/Fe] of CS~22963--004
is estimated with UV OH lines \citep{lai08}. The O abundance derived from
these lines could be reduced by the 3D effect as much as $0.7$
\citep[\eg][]{asp05,gon10}. The SN model is consistent with the
observation, even if [O/Fe] is reduced by
$0.7$~dex.

{\bf CS~22968--014}: The metallicity is [Fe/H]~$=-3.58$ and the
abundance ratios are shown in Figure~\ref{fig:EMP1}g
\citep{cay04,spi05,spi06,and07,and08,and10,spi12,yon13}. The
abundance pattern except for [K/Fe] is well reproduced by an
SN model with $\E=20$,
$\Mbh=3.84\Msun$, and $\Mfe=8.61\times10^{-2}\Msun$. 
The improvement of [K/Fe]
could be obtained with including the $p$-rich ejecta produced by neutrino
processes (see also a discussion on CS~22189--009). The abundance
pattern and thus the SN model are similar to CS~22189--009. 

{\bf $\star$ [O/Fe]}: \cite{cay04} presented
[O/Fe] tentatively corrected by the 3D effect which is computed
by following \cite{nis02}. The corrected [O/Fe] is in good agreement with the SN
model.

{\bf HE~1356--0622}: The metallicity is [Fe/H]~$=-3.58$ and the
abundance ratios are shown in Figure~\ref{fig:EMP1}h \citep{coh08,yon13}.
For CNO elements, only an upper limit to [C/Fe] is obtained as $+0.2$,
which is rather low compared with [Na/Fe]$\sim0.6$. 
The abundance pattern except for [Na/Fe] is well reproduced by an SN model with
$\E=5$, $\Mbh=1.96\Msun$, and $\Mfe=1.11\times10^{-1}\Msun$. 
The high [(Sc,~Ti)/Fe] and the upper limit
to [Zn/Fe] are reproduced with large $\MCini$.

{\bf $\star$ [Na/C]}: The upper limit of [C/Fe] makes it difficult to reproduce high
[Na/Fe]. Such high [Na/C] ($>+0.4$) could be reproduced if C is converted to N by
mixing in HE~1356--0622 as in the NEMP stars (\S~\ref{sec:NEMP}). Also, the
underproduction of Na could be improved with a lower explosion energy
because weak explosive burning suppresses the destruction of
Na. However, such an explosion with $\E\leq1$ is difficult to
reproduce [(Sc,~Ti)/Fe].

{\bf HE~1347--1025}: The metallicity is [Fe/H]~$=-3.62$ and the
abundance ratios are shown in Figure~\ref{fig:EMP1}i \citep{coh08,yon13}.
The abundance pattern except for [Sc/Fe] is well reproduced by an SN model with $\E=5$,
$\Mbh=4.00\Msun$, and $\Mfe=9.85\times10^{-2}\Msun$. 

{\bf $\star$ [Sc/Fe] and [Ti/Fe]}: 
The overproduction of Sc could be
suppressed by a model with slightly lower entropy but such a model would
lead to the underproduction of Ti.

{\bf BS~16076--006}: The metallicity is [Fe/H]~$=-3.62$ and the
abundance ratios are shown in Figure~\ref{fig:EMP1}j \citep{bon09,yon13}.
The star shows [C/N]~$\sim0$. Since the star is a cool subgiant
\citep{bon07}, the observed N is likely to be
synthesized in the SN progenitor due to the enhanced mixing. The abundance
pattern is well reproduced by an SN model with $\E=5$, $\Mbh=3.87\Msun$,
and $\Mfe=5.10\times10^{-2}\Msun$. 

{\bf CS~22876--032AB}: The abundance ratios are presented in
\cite{gon08}. These two stars with similar masses compose a binary system.
The abundance patterns of these stars are consistent within the errors,
except for Li, and their metallicities are [Fe/H]~$=-3.66$
(the red and blue circles in Fig.~\ref{fig:EMP2}a). The abundance
patterns are well reproduced by an SN
model with $\E=20$, $\Mbh=4.49\Msun$, and
$\Mfe=4.24\times10^{-2}\Msun$. 

{\bf $\star$ [O/Fe]}: 
The 3D effect largely
reduces [O/Fe] derived from the UV OH lines. If the 3D correction
would be negligible, the high 1D [O/Fe] indicates an SN with small
$\Mfe$ that leads to high [C/Fe] (see also \S~\ref{sec:CEMP}).

{\bf HE~0049--3948}: The metallicity is [Fe/H]~$=-3.62$ and the
abundance ratios are shown in Figure~\ref{fig:EMP2}b \citep{yon13}.
For CNO elements, only loose upper limits to [(C,~N)/Fe] are
obtained. The abundance pattern is well
reproduced by an SN model with $\E=20$, $\Mbh=4.28\Msun$,
and $\Mfe=8.34\times10^{-2}\Msun$.
The slightly high [Ni/Fe] is reproduced by ejection of material with low $\ye$
($<0.5$).

{\bf HE~1320--2952}: The metallicity is [Fe/H]~$=-3.69$ and the
abundance ratios are shown in Figure~\ref{fig:EMP2}c \citep{yon13}. 
For CNO elements, marginal upper limits to [(C,~N)/Fe] are obtained.
The abundance pattern is well
reproduced by an SN model with $\E=10$, $\Mbh=2.73\Msun$,
and $\Mfe=1.30\times10^{-1}\Msun$.

{\bf BS~16467--062}: The metallicity is [Fe/H]~$=-3.74$ and the
abundance ratios are shown in Figure~\ref{fig:EMP2}d
\citep{cay04,spi05,spi06,and07,and08,lai08,coh08,and10,spi12,yon13}.
The abundance pattern except for [(K,~V,~Ni)/Fe] is well reproduced by an
SN model with $\E=20$, $\Mbh=4.51\Msun$, and
$\Mfe=4.36\times10^{-2}\Msun$ (solid line). A model with the lower explosion
energy of $\E=5$ also well reproduces the abundance pattern (dashed
line). It is
difficult to distinguish the SN models with different explosion
energies for this star (\S~\ref{sec:energy}). 

{\bf $\star$ [(V,~Ni)/Fe]}: The abundance ratios
[(V,~Ni)/Fe] depend on $\ye$ and the mass of the complete Si-burning
layer. In complete Si burning, since nuclei are photodisintegrated and 
re-synthesized, the isotopic ratios depend mainly on $\ye$. The synthesis
of V and Ni is efficient in the regions where $\ye$ is close to $\ye$ of
the parent radioisotopes, \eg $^{51}$Mn for V and $^{58}$Cu for
Ni, or $\ye$ of the stable isotopes, \eg $^{58}$Ni for Ni.
Since $\ye$ also affects the abundances of other Fe-peak elements, the abundance ratios
among the Fe-peak elements could constrain $\ye$
distribution in the SN ejecta. A multi-dimensional effect also
influences the synthesis of these elements.

{\bf CS~22172--002}: The metallicity is [Fe/H]~$=-3.74$ and the
abundance ratios are shown in Figure~\ref{fig:EMP2}e
\citep{cay04,spi05,spi06,and07,and08,and10,spi12,yon13}.
The abundance pattern except for [K/Fe] is well reproduced by an SN model with
$\E=20$, $\Mbh=3.87\Msun$, and $\Mfe=7.76\times10^{-2}\Msun$. 
The abundance pattern and thus the SN model are similar to
CS~22189--009. 

{\bf HE~0228--4047}: The metallicity is [Fe/H]~$=-3.75$ and the
abundance ratios are shown in Figure~\ref{fig:EMP2}f \citep{yon13}. The
abundance patterns derived with the surface gravity of subgiant and dwarf
stars are similar. For CNO elements, only a loose upper limit to [C/Fe]
is obtained. The abundance patterns are well reproduced
by an SN model with $\E=20$, $\Mbh=3.27\Msun$,
and $\Mfe=1.36\times10^{-1}\Msun$. 
The high [Ni/Fe] is
reproduced by ejection of low-$\ye$ material and the high [Ti/Fe]
indicates nucleosynthesis in a high-entropy environment. 

{\bf G~238-30}: The metallicity is [Fe/H]~$=-3.77$ and the
abundance ratios are shown in Figure~\ref{fig:EMP2}g \citep{ste02,ish10}.
The abundance ratios [(C,~N,~O)/Fe] are not available. The
abundance pattern is well reproduced by an SN model with $\E=5$,
$\Mbh=2.55\Msun$, and $\Mfe=6.80\times10^{-2}\Msun$.

{\bf CS~22885--096}: The metallicity is [Fe/H]~$=-3.77$ and the
abundance ratios are shown in Figure~\ref{fig:EMP2}h
\citep{cay04,spi05,spi06,and07,and08,coh08,and10,spi12,yon13}.
Since the star shows [C/N]~$\sim0$ and the mixing in CS~22885--096 is not
strong \citep{spi05,spi06}, N is likely to be synthesized in the SN
progenitor due to the enhanced mixing. The abundance pattern except for
[K/Fe] is well reproduced by an SN model with $\E=20$, $\Mbh=3.76\Msun$,
and $\Mfe=4.95\times10^{-2}\Msun$. The abundance pattern and thus the SN model are
similar to those of CS~22189--009.

\subsection{Carbon-Enhanced Metal-Poor stars}
\label{sec:CEMP}

The CEMP star is one of the largest subclasses of the MP stars.
They are further classified by the abundance of $n$-capture elements:
CEMP-s stars with an enhancement of $s$-process elements, CEMP-no stars
without any enhancement of $n$-capture elements, and so on
(see \citealt{bee05} for detail). 

The CEMP-s star is most common among CEMP stars and is considered to be
originated from an MP star constituting a binary system. A
mass transfer from an asymptotic giant branch (AGB) binary companion enhances
abundances of carbon and $s$-process elements on the surface 
of the MP star and transforms the MP star to a CEMP-s
star \citep[e.g.,][]{bus99,iwa04,sud04,lug12}. A signature of
binarity is seen for possibly all CEMP-s stars \citep{luc05}. The lowest
[Fe/H] of CEMP-s stars is $-3.1$ \citep{joh02}.

We focus on 12 CEMP-no stars with $-4.1\lsim$~[Fe/H]~$\lsim-3.5$,
[C/Fe]~$>+1.0$, and [C/N]~$\geq-0.5$. In contrast to the origin of
CEMP-s stars, the origin of CEMP-no stars is under debate \citep{nom13}. The mechanism
of carbon enhancement has been proposed as follows:
\begin{description}
\item[(1)] a faint SN with small ejection of Fe
\citep{ume02a,ume05a,tom07b},\footnote{The SN with small ejection of
Fe is faint because Fe is synthesized as \Nifs\ that powers a light
curve of the SN.}
\item[(2)] a dark SN without Fe ejection \citep{lim03},
\item[(3)] mass loss from a rotating massive star \citep{mey06},
\item[(4)] a mass transfer from an AGB binary companion \citep[e.g.,][]{sud04},
	   and 
\item[(5)] self-enrichment in an MP red giant star
\citep{fuj00,cam10}. 
\end{description} 

Among the above 5 models, only the faint SN model attributes the origin
of the CEMP-no stars to a {\sl single} Pop III SN. Other models explain
the carbon enhancement but require another SN(e) to produce elements
heavier than CNO and/or Fe-peak elements. In the latter models, Fe-peak
elements are suggested to originate from 
concurrent enrichment by a normal SN or hypernova
\citep{lim03} and/or the mass accretion from the interstellar medium
\citep{yos81}. The mass accretion from the interstellar medium could
also produce the CEMP star only by itself if the interstellar medium is
C-rich.

We adopt a faint SN model as the mechanism of carbon
enhancement for the following reasons: 
\begin{description}
\item[(a)] A CEMP-no star, which can be
explained only by the faint SN mechanism, has been discovered \citep{ito09,ito13}.
\item[(b)] Faint SNe are observed in the
present day [$\Mni\sim10^{-3}\Msun$, \eg SN~1999br:
\citealt{zam03}, SN~2008ha: \citealt{moriya10}]. 
\item[(c)] Concurrent and equivalent enrichment by two
	   SNe or massive stars may be rare.
\item[(d)] The amount of mass loss from a rapidly rotating
$Z=0$ star is small \citep{eks08}. 
\item[(e)] The CEMP-no stars show low [Ba/Fe].\footnote{We note that the abundance of
	   Ba could be suppressed if the
	   binary companion is the low-mass
	   AGB star with $\Mms<1.2\Msun$ or massive AGB star with $\Mms>3.5\Msun$
	   \citep{kom07,sud10}.}
\item[(f)] The binary fraction of the CEMP-no stars is
absolutely different from that of the CEMP-s stars \citep{nor13p4}. 
\item[(g)] There exist CEMP-no subgiants and dwarf stars and thus the
	   self-enrichment mechanism is not applicable for all CEMP-no stars. 
\end{description} 

{\bf HE~1150--0428}: The metallicity is [Fe/H]~$=-3.47$ and the
abundance ratios are shown in Figure~\ref{fig:CEMP}a
\citep{coh06,yon13}. 

{\bf $\star$ [C/N] and [(B,~F)/Fe]}: Since N is as enhanced as [C/N]~$\sim0$, origin of
N needs to be discussed. If N is assumed to be synthesized in the SN
progenitor by the enhanced mixing of H into the He layer and the
subsequent CNO cycle, the abundance ratios [(C,~N,~Mg,~Ti)/Fe] are
well reproduced by an SN model with $\E=5$, $\Mbh=5.83\Msun$,
and $\Mfe=3.43\times10^{-4}\Msun$ (solid line). However, HE~1150--0428
shows low $^{12}$C/$^{13}$C ($\sim4$, \citealt{coh06}) in spite of
the relatively high surface gravity [$\log(g)=2.54$]. This indicates
that a material in the envelope experiences the CNO cycle, and thus suggests
that N is produced in HE~1150--0428 as in the NEMP stars
(\S~\ref{sec:NEMP}). If N is assumed to be converted from C by
the CNO cycle in HE~1150--0428, the abundance ratios are well
reproduced by an SN model of Model A with $\E=5$, $\Mbh=5.83\Msun$,
and $\Mfe=3.51\times10^{-4}\Msun$ (dashed line).
The origin of N might be
constrained by [(B,~F)/Fe] because B and F are synthesized by the hot CNO
cycle and $^3$He burning if the He/N-rich layer experiences high temperature during the
shock propagation. 

{\bf $\star$ [C/Mg]}: The high [C/Mg] is
reproduced by {\sl mixing-and-fallback} during the explosion: mixing up
to the top of the O+Mg layer, \ie large
$\Mmout$, and the small ejection from the mixing region, \ie small
$\fej$. These result in large $\Mbh$. 

{\bf $\star$ [Ca/Fe]}: There remains a discrepancy on
[Ca/Fe]. The large $\MCini$ could enhance [Ca/Fe] but lead to too high
[Mg/Fe] and too low [Ti/Fe].

{\bf HE~1012--1540}:  The metallicity is [Fe/H]~$=-3.48$ and the
abundance ratios are shown in Figure~\ref{fig:CEMP}b
\citep{coh08,yon13}. The abundance pattern except for [N/Fe] is
well reproduced by an SN model with $\E=20$, $\Mbh=4.81\Msun$,
and $\Mfe=1.11\times10^{-3}\Msun$. 

{\bf $\star$ [(C,~Mg)/Fe]}: The high [(C,~Mg)/Fe] is
reproduced by the mixing-and-fallback model that assumes mixing up to the
middle of the O+Mg layer and small ejection from the mixing region.

{\bf $\star$ [C/N]}: The N abundance is as low as
[N/(C,~O)]~$\lsim-1.0$ and thus the faint SN mechanism is most preferable
among the C enhancement mechanisms. The underproduction of N is
improved if $15\%$ of C in the He layer is converted to N due to
slightly efficient mixing at the boundary of H and He layers in
the SN progenitor.  

{\bf CS~29498--043}: The metallicity is [Fe/H]~$=-3.52$ and the
abundance ratios are shown in Figure~\ref{fig:CEMP}c
\citep{aok04,yon13}. The abundance pattern is well reproduced by the
SN model A with $\E=20$, $\Mbh=5.11\Msun$, and
$\Mfe=9.08\times10^{-4}\Msun$ (solid line).

{\bf $\star$ [C/N]}: The N abundance is as high as [N/C] $\sim$ 0, but
the origin of N cannot be constrained, since there is no observation of
Li or $^{13}$C. The abundance pattern is also well
reproduced with the SN model B if that the mixing
in CS~29498--043 is assumed to enhance [N/Fe] as in HE~1150--0428
(dashed line);
however, it needs to be confirmed
whether [C/N]~$\sim0$ is realized. The differences between
Models A and B appear in [(B,~F)/Fe].

{\bf HE~1506--0113}: The metallicity is [Fe/H]~$=-3.54$ and the
abundance ratios are shown in Figure~\ref{fig:CEMP}d \citep{yon13}.
The abundance pattern except for [(N,~Na,~Ni)/Fe] is well reproduced by
an SN model with $\E=20$,
$\Mbh=5.15\Msun$, and $\Mfe=6.80\times10^{-3}\Msun$. Although the NLTE abundance
determination reduces [Na/Fe], the underproduction of Na is not
solved. The agreement of [Ni/Fe] is improved if low-$\ye$
material is ejected.

{\bf $\star$ [C/N]}: The N abundance is as low as
[N/C]~$\lsim-0.9$ and thus the faint SN mechanism is favored among the C enhancement
mechanisms. [N/Fe] could be reproduced if
$15\%$ of C in the He layer is converted to N due
to slightly efficient mixing at the boundary of H and He layers in
the SN progenitor. 

{\bf CS~29527--015}: The metallicity is [Fe/H]~$=-3.55$ and the
abundance ratios are shown in Figure~\ref{fig:CEMP}e
\citep{and07,and08,bon09,and10,spi12,yon13}. The abundance pattern
except for [Na/Fe] is
well reproduced by an SN model with $\E=20$, $\Mbh=5.14\Msun$,
and $\Mfe=6.52\times10^{-3}\Msun$.

{\bf $\star$ [Na/Mg]}: The observed low [Na/Mg] is difficult to
be  reproduced unless the Na/Mg ratio in the O+Mg layer of the progenitor
star is reduced by, \eg quasi-static C burning at higher temperature.
Thus, the agreement is slightly improved with the
$40\Msun$ model (\S~\ref{sec:40Msun}). 

{\bf HE~0132--2429}: The metallicity is [Fe/H]~$=-3.70$ and the
abundance ratios are shown in Figure~\ref{fig:CEMP}f
\citep{coh08,yon13}. The abundance pattern except for [(Na,~V,~Cu)/Fe]
is well reproduced by an SN model with
$\E=20$, $\Mbh=5.39\Msun$, and $\Mfe=1.24\times10^{-2}\Msun$.

{\bf $\star$ [C/N]}: The observed [N/C] is as high as $\sim0.2$. In the
above model, we assume that N is synthesized by the enhanced mixing in
the supernova progenitor, since there is no constraint on the origin of
enhanced N and it is non-trivial whether [C/N]$\sim0$ is realized by the
mixing in HE~0132--2429. 

{\bf $\star$ [Cu/Fe]}: As mentioned for
CS~22963--004, small $\MCini$ can yield low [Cu/Fe] but would lead
to the reduction of [(Sc,~Ti,~V)/Fe]. 

{\bf $\star$ [Na/Fe]}: The disagreement of [Na/Fe] could be improved
if the convective shell has higher temperature during quasi-static C burning.

{\bf HE~1300+0157}: The metallicity is [Fe/H]~$=-3.76$ and the
abundance ratios are shown in Figure~\ref{fig:CEMP}g
\citep{bar05,fre07,bee07,coh08,yon13}. Among the many papers
working on HE~1300+0157, we construct an SN model
to reproduce the abundance ratios in \cite{fre07} because
they intensively studied HE~1300+0157. 

The abundance pattern for surface
gravity of a subgiant star is
well reproduced by the SN model A (solid line) with $\E=20$, $\Mbh=3.25\Msun$,
and $\Mfe=5.21\times10^{-2}\Msun$, 
except for [(Na,~Cu)/Fe]. On the other hand, the abundance pattern for
surface gravity of a dwarf star is reproduced by an SN model without
{\sl mixing-and-fallback}. The properties of the 
SN model B (dashed line) are $\E=20$, $\Mbh=2.16\Msun$,
and $\Mfe=3.42\times10^{-1}\Msun$.
We also present an SN model C for the abundance
patterns without 3D effects on [(C,~O)/Fe]. The properties are $\E=20$,
$\Mbh=5.57\Msun$, and $\Mfe=3.58\times10^{-3}\Msun$ (dotted line).

{\bf $\star$ [C/Fe]}: Since [C/Fe] for the dwarf surface gravity (model
B) is lower than
that for the subgiant surface gravity (model A), the resultant $\Mfe$
for model B is $\sim7$ times larger than for model A. The discrimination
between models A and B is important to constrain $\Mfe$. 

{\bf BD+44$^\circ$493}: The metallicity is [Fe/H]~$=-3.78$ and the
abundance ratios are shown in Figure~\ref{fig:CEMP}h
\citep{ito09,ito13}.\footnote{The progenitor models adopted in the
present paper are different from \cite{ito13} in order to investigate the
effect of the enhanced mixing.} 
The abundance pattern except for [(N,~Na,~Ti,~Cu)/Fe] is well reproduced by an SN model with $\E=5$,
$\Mbh=5.33\Msun$, and $\Mfe=6.17\times10^{-3}\Msun$. The
discrepancies on [(Na,~Ti,~Cu)/Fe] are improved with the $40\Msun$ model
(\S~\ref{sec:40Msun}).

{\bf $\star$ [(C, N)/Fe]}: 
The origin of the C enhancement in
BD+44$^\circ$493 is clearly constrained to be a faint SN by virtue of
high [C/N] ($>0$), low [C/O] ($<0$), and low Ba and Pb abundance
([Ba/Fe]$=-0.6$ and [Pb/Fe]$<+1.98$, \citealt{ito13}).
[N/Fe] could be
reproduced if $10\%$ of C in the He layer is converted to N due to
slightly enhanced mixing in the SN progenitor. 

{\bf HE~1201--1512}: The metallicity is [Fe/H]~$=-3.89$ and the
abundance ratios are shown in Figure~\ref{fig:CEMP}i \citep{yon13}. 
\cite{yon13} provides two kinds of abundance patterns with surface
gravity of dwarf and subgiant stars. The abundance pattern for 
the dwarf surface gravity is well reproduced by an SN model
with $\E=5$, $\Mbh=5.65\Msun$, and
$\Mfe=9.17\times10^{-3}\Msun$, 
except for [Co/Fe] (solid line). The abundance pattern for the subgiant surface
gravity is reproduced by an SN model with $\E=5$, $\Mbh=5.72\Msun$,
and $\Mfe=4.53\times10^{-3}\Msun$ (dashed line).
The high [Co/Fe] is difficult to be reproduced
without an inhomogeneous mixing in the SN ejecta (see also a discussion
on BS~16545--089 in \S~\ref{sec:EMP}).

{\bf HE~2139--5432}: The metallicity is [Fe/H]~$=-4.02$ and the
abundance ratios are shown in Figure~\ref{fig:CEMP}j
\citep{yon13}. The
abundance pattern except for [Si/Fe] is well reproduced by an SN model with $\E=5$,
$\Mbh=5.03\Msun$, and $\Mfe=6.78\times10^{-4}\Msun$.

{\bf $\star$ [C/N]}: 
Although there is no constraint on the origin of N enhancement, we
assume that N is synthesized by the enhanced mixing in the progenitor
because of [C/N]~$\sim0$ as in HE~0132--2429. 

{\bf $\star$ [Si/Fe]}: 
[Si/Fe] could be enhanced by a large
$\MCini$ but such a large $\MCini$ would reduce [Ti/Fe].

\begin{figure*}
\epsscale{.9}
\plotone{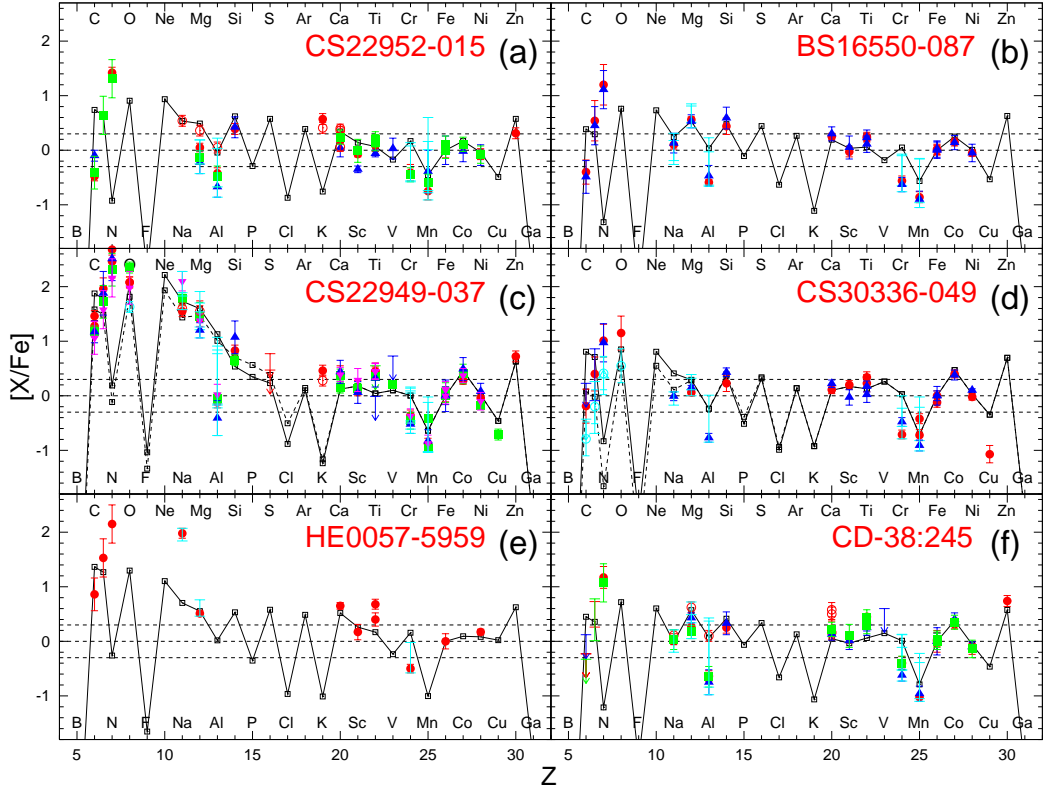}
\figcaption{Same as Fig.~\ref{fig:EMP1} but for the NEMP stars. \label{fig:NEMP}}
\end{figure*}

{\bf 53327-2044-515}:  The metallicity is [Fe/H]~$=-4.04$ and the
abundance ratios are shown in Figure~\ref{fig:CEMP}k
\citep{yon13,nor13p4}. The abundance pattern for the dwarf surface
gravity is
well reproduced by an SN model with $\E=5$, $\Mbh=5.56\Msun$,
and $\Mfe=8.51\times10^{-3}\Msun$, 
except for [Ti/Fe] (solid line). 
Assuming 53327-2044-515 is a subgiant star, the abundance pattern
including [Ti/Fe] is well
reproduced by an SN model with $\E=5$, $\Mbh=5.66\Msun$,
and $\Mfe=3.26\times10^{-3}\Msun$ (dashed line).

{\bf G~77--61}:  The metallicity is [Fe/H]~$=-4.08$ and the
abundance ratios are shown in Figure~\ref{fig:CEMP}l
\citep{ple05,bee07}. Since its surface gravity is
$\log(g)=5.05$, N is likely to be synthesized by the enhanced
mixing in the progenitor. The abundance pattern except for [K/Fe] is well reproduced by an
SN model with $\E=1$, $\Mbh=6.17\Msun$, and $\Mfe=1.85\times10^{-4}\Msun$.
The high [(C,~N,~O)/Mg] and [(Ca,~Cr)/Fe] are reproduced by a large
$\Mmout$ and small $\fej$, and a large $\MCini$, respectively.

{\bf $\star$ [K/Fe]}: 
[K/Fe] of G~77--61 is remarkably
high. Explosive nucleosynthesis in the $p$-rich ejecta
might explain such high K abundance, but could also yield a large amount of odd
elements \citep{iwa06}. According to \cite{iwa06}, overabundance of
P,~Cl,~Sc,~V ([X/Fe]~$>+0.5$) is predicted. 

{\bf $\star$ [(Sc,~Ti,~Co,~Zn)/Fe]}: The lack of observed
[(Sc,~Ti,~Co,~Zn)/Fe] makes it difficult to constrain the explosion
energy and entropy during explosive burning.

\subsection{Nitrogen-Enhanced Metal-Poor stars}
\label{sec:NEMP}

Among the EMP stars, there are stars that show enhancement of N, called
NEMP stars, which are defined by
[N/Fe]~$>0.5$ and [C/N]~$<-0.5$ \citep[\eg][]{joh07}. Half of the NEMP
stars at [Fe/H]~$\lsim-3.0$ display a signature of mixing in the Li
abundance and the $^{13}$C/$^{12}$C ratio \citep{spi05,spi06}, and the others do
not have these measurement due to observational difficulty. 

The stars that undergo extra mixing on the red giant branch (RGB) in
addition to the first dredge-up are called ``mixed'' stars. The mixing
in the MP stars enhances the
conversion from C to N via the CNO cycle. This could be a dominant process
to create the NEMP stars at low metallicity. Thus, although
the Li abundance or $^{13}$C/$^{12}$C ratio are not available for some
of the NEMP stars, we assume the NEMP stars at [Fe/H]$\lsim-3.5$ are
``mixed'' stars and construct an SN model that reproduces [(C+N)/Fe]
instead of individual [C/Fe] and [N/Fe].

If the mixing on the RGB is inefficient in the NEMP stars, N should be enhanced by
other mechanisms. The mixing in the AGB
stars could enhance [N/C] up to $\sim+2$ \citep{nis09,kar10} and the
mass accretion from the AGB binary companion could transform an MP
star to an NEMP star \citep[\eg][]{pol12}. However, binarity has been
found only for
BS~16550--087 \citep{lai08}. 

On the other hand, mixing in a massive SN progenitor can enhance N only up to [N/C]~$\sim+0.5$
\citep{iwa05,mey06}. Thus the process is not appropriate, by
definition, for the origin of the NEMP stars. 

\begin{figure*}
\epsscale{.9}
\plotone{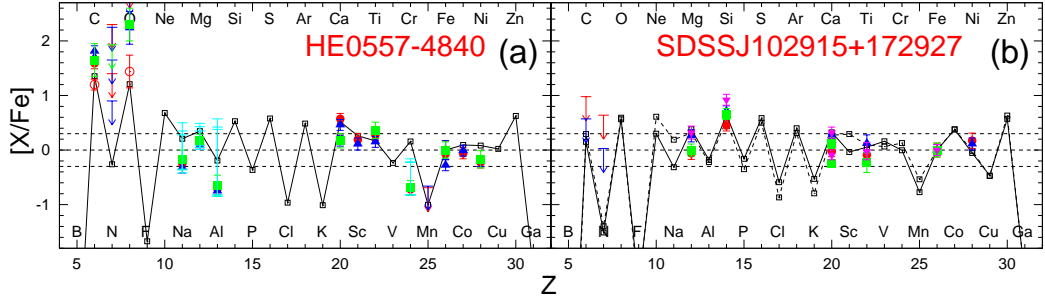}
\figcaption{Same as Fig.~\ref{fig:EMP1} but for the UMP stars.\label{fig:UMP}}
\end{figure*}

{\bf CS~22952--015}: The metallicity is [Fe/H]~$=-3.44$ and the
abundance ratios are shown in Figure~\ref{fig:NEMP}a
\citep{cay04,hon04,spi05,spi06,and07,and08,and10,spi12,yon13}.
CS~22952--015 is a mixed star \citep{spi05,spi06} and thus N is likely
to be converted from C in CS~22952--015. The abundance pattern except
for [K/Fe] is well reproduced by an SN model with $\E=5$, $\Mbh=4.46\Msun$,
and $\Mfe=3.09\times10^{-2}\Msun$. The model demonstrates that
[C/Fe] of CS~22952--015 before the conversion from C to N is less
than $+1.0$ and that CS~22952--015 was the EMP star at
birth.

{\bf BS~16550--087}: The metallicity is [Fe/H]~$=-3.54$. The
abundance ratios are shown in Figure~\ref{fig:NEMP}b
\citep{lai08,yon13} but there is no observation of Li or
$^{13}$C. The abundance pattern is well reproduced by an SN model with
$\E=10$, $\Mbh=3.14\Msun$, and
$\Mfe=7.68\times10^{-2}\Msun$. According to the SN model,
BS~16550--087 was the EMP star at birth.

{\bf CS~22949--037}: The metallicity is [Fe/H]~$=-3.89$ and the
abundance ratios are shown in Figure~\ref{fig:NEMP}c
\citep{nor01,nor02,cay04,spi05,spi06,and07,coh08,and10,spi11,spi12,yon13}.
CS~22949--037 is a mixed star \citep{spi05,spi06}. The abundance pattern
except for [K/Fe] is well
reproduced by an SN model with $\E=5$, $\Mbh=4.34\Msun$,
and $\Mfe=2.42\times10^{-3}\Msun$ (solid line). 
As mentioned for
CS~22189--009 in \S~\ref{sec:EMP}, the underproduction of K could be
solved by including
$p$-rich ejecta due to neutrino processes \citep[\eg][]{iwa06,heg10}.
The model demonstrates that CS~22949--037 was the CEMP star at birth. 

{\bf $\star$ [O/Fe]}: 
\cite{cay04} also showed [O/Fe] tentatively corrected by the 3D effect. 
The corrected [O/Fe] is reproduced by
an SN model with $\E=10$, $\Mbh=4.11\Msun$, and
$\Mfe=4.85\times10^{-3}\Msun$ (dashed line). The SN model has two times larger
$\Mfe$ than the SN model for the uncorrected [O/Fe]. 

\cite{pol12} suggests that
CS~22949--037 is originated by a mass transfer from an N-rich AGB
companion because of insufficient time to alter the surface
abundance by extra mixing \citep{sta09}. However, there is no
indication of the binarity of CS~22949--037 \citep{dep02}. Furthermore,
it is not clearly shown whether the abundance pattern of CS~22949--037,
\eg high [(Na,~Al)/Mg], is reproduced by this mechanism.

{\bf CS~30336--049}: The metallicity is [Fe/H]~$=-4.07$. The
abundance ratios are shown in Figure~\ref{fig:NEMP}d
\citep{lai08,yon13}. The abundance pattern
except for [(Na,~Cu)/Fe] is well reproduced by an SN model with $\E=5$, $\Mbh=5.01\Msun$,
and $\Mfe=2.54\times10^{-2}\Msun$ (solid line).

{\bf $\star$ [C/N]}: 
The low [C/N] ($\sim-1.0$) cannot be realized in
the mixing in the massive SN progenitor, while there is no observation of Li
or $^{13}$C. 

{\bf $\star$ [(C,~N,~O)/Fe] and [Na/Fe]}: 
The CNO abundance of CS~30336-049 is estimated with CH, NH, OH lines
suffered from the large 3D effects (\eg \citealt{asp05}). The
overproduction of Na could be resolved if [(C,~N,~O)/Fe] is reduced by
$0.6$~dex due to the 3D effects (dashed line).

{\bf HE~0057--5959}: The metallicity is [Fe/H]~$=-4.08$. The
abundance ratios are shown in Figure~\ref{fig:NEMP}e
\citep{nor13p4,yon13} but there is no observation of $^{13}$C. The
abundance pattern except for [Na/Fe] is well reproduced by an SN model with
$\E=5$, $\Mbh=5.38\Msun$, and $\Mfe=6.80\times10^{-3}\Msun$. 

{\bf $\star$ [Na/Mg]}: 
The high [Na/Mg] requires similar
number fractions of Na and Mg at the O+Mg layer in the progenitor.
The disagreement of [Na/Mg], the surface gravity of $\log(g)=2.65$, and
the high Li abundance ($\log\epsilon({\rm Li})=2.12$, \citealt{nor13p4})
might imply that HE~0057--5959 is originated from the mass transfer from
an N-rich AGB companion. However, the binarity is not well constrained
and the star shows low [Ba/Fe] ($=-0.46$, \citealt{yon13}).

{\bf CD--38:245}: The metallicity is [Fe/H]~$=-4.11$ and the
abundance ratios are shown in Figure~\ref{fig:NEMP}f
\citep{nor01,cay04,spi06,and07,and08,and10,spi12,yon13}. CD--38:245 is a
mixed star \citep{spi05,spi06}. Since CD--38:245 has only an upper limit to [C/Fe],
a range of [(C+N)/Fe] is estimated by assuming the range of [C/Fe] from $-\infty$ to $-0.28$. 
The abundance pattern is well reproduced by an SN model with $\E=20$,
$\Mbh=4.07\Msun$, and $\Mfe=6.10\times10^{-2}\Msun$.

\subsection{Ultra Metal-Poor stars}
\label{sec:UMP}

The UMP stars are defined by the metallicity of [Fe/H]~$<-4.0$
\citep{bee05}. In this subsection, we focus on two UMP stars with
[Fe/H]~$<-4.5$: HE~0557--4840 \citep{nor07} and
SDSS~J102915+172927 \citep{caf11nat}. The other UMP stars with
[Fe/H]~$\gsim-4.1$ are shown in \S~\ref{sec:CEMP} and \S~\ref{sec:NEMP}.

{\bf HE~0557--4840}: The metallicity is [Fe/H]~$=-4.77$ and the
abundance ratios are shown in Figure~\ref{fig:UMP}a
\citep{nor07,nor12,yon13}. The abundance pattern except for [Cr/Fe] is well reproduced by an
SN model with $\E=5$, $\Mbh=5.57\Msun$, and $\Mfe=6.80\times10^{-3}\Msun$.

{\bf $\star$ [Cr/Fe]}: 
[Cr/Fe] is
overproduced even with taking into account the NLTE correction of
$+0.4$~dex adopted in this paper. However, the NLTE correction for
[Cr/Fe] could be larger at lower [Fe/H] \citep{ber10cr} and improve the agreement.

\begin{figure*}
\epsscale{.9}
\plotone{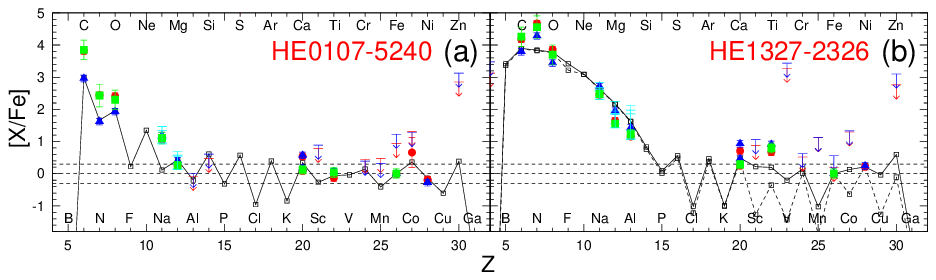}
\figcaption{Same as Fig.~\ref{fig:EMP1} but for the HMP stars.\label{fig:HMP}}
\end{figure*}

{\bf SDSS~J102915+172927}: The metallicity is [Fe/H]~$=-4.92$ and the
abundance ratios are shown in Figure~\ref{fig:UMP}b
\citep{caf11nat,caf12}. This star is the most ``metal''-deficient stars
discovered so far. The abundance pattern is well reproduced by an
SN model
with $\E=20$, $\Mbh=3.03\Msun$, and $\Mfe=1.21\times10^{-1}\Msun$
(solid line). However, the number of measured
abundance ratios are rather few. Thus the abundance pattern
can be reproduced by a lower $E$ model with $\E=5$, $\Mbh=3.49\Msun$,
and $\Mfe=9.25\times10^{-2}\Msun$ as well (dashed line).

Although [Fe/H] of the UMP stars are in between the EMP and HMP stars,
metal abundance ratios of the UMP
stars are similar to those of the EMP or CEMP stars. This indicates that
the parent SNe are similar to those of the EMP or CEMP stars. 

[Fe/H] of EMP stars depends not only on the Fe mass
ejected by an SN but also on the H mass which mixes with the SN
ejecta. Although the H mass is determined mainly by the explosion energy
in the 1D SN-induced star formation \citep{shi98,tho98},
sophisticated 3D simulations show that [Fe/H] of a mixture
gas are widely distributed from $-5$ to $-1$ with an average similar to
the 1D calculations \citep{naks00,rit12}. 
Therefore, we suggest that the
UMP stars form from a cloud with lower [Fe/H] which was enriched by the
SN whose property was similar to SNe to originate the EMP or CEMP stars;
this suggests that the EMP or CEMP stars also form around the same SN.

\subsection{Hyper Metal-Poor stars}
\label{sec:HMP}

The most Fe-deficient stars are two HMP stars, HE~0107--5240
and HE~1327--2326. The HMP stars show similarly high [C/Fe] but
large differences in [C/(N,~O)] and [(Na,~Mg,~Al)/Fe]. The 5 mechanisms to
enhance [C/Fe] have been suggested and still under
debate (\S~\ref{sec:CEMP}). Here, we adopt the faint SN mechanism which
successfully explains the similarities and differences in the HMP stars
\citep{ume03,iwa05,tom07a}.

{\bf HE~0107--5240}: The metallicity is [Fe/H]~$=-5.61$ and the
abundance ratios are shown in Figure~\ref{fig:HMP}a
\citep{chr02,chr04,bes05,col06,yon13}. The abundance pattern
except for [Na/Fe] is well reproduced by an SN model with $\E=5$,
$\Mbh=6.24\Msun$, and $\Mfe=8.02\times10^{-5}\Msun$.

{\bf $\star$ [C/Fe]}: 
Since the 3D effect considerably reduces [C/Fe], the ejected Fe mass is increased by a
factor of $\sim5$ compared to the SN model in \cite{iwa05}.

{\bf $\star$ [Co/Fe]}: 
Although \cite{ume03} and \cite{iwa05} attributed the abundance
pattern of HE~0107--5240 to an SN with $E\leq10^{51}$~ergs, such a
low-$E$ explosion cannot reproduce high [Co/Fe] of HE~0107--5240
\citep{bes05}. Thus, assuming that large fallback takes place in an
aspherical explosion \citep{tom07a}, we adopt an energetic explosion
model in which [Co/Fe] is enhanced due to explosive nucleosynthesis in
the high-entropy environment \citep{tom09a}. 

{\bf $\star$ [Na/Fe]}: 
The agreement of [Na/Fe] could be improved with taking
into account
the NLTE effect, although the naive extrapolation of \cite{and07} enhances
[Na/Fe]. 

{\bf HE~1327--2326}: The metallicity is [Fe/H]~$=-5.88$ and the
abundance ratios are shown in Figure~\ref{fig:HMP}b
\citep{fre05,fre08,col06,bon12,yon13}. Some mixing processes might take
place in HE~1327--2326 since Li is not observed. Nevertheless, we assume
that N is synthesized by the enhanced mixing in the SN progenitor because
HE~1327--2326 is a subgiant star. The abundance pattern is well
reproduced by an SN
model with $\E=5$, $\Mbh=5.72\Msun$, and $\Mfe=1.45\times10^{-5}\Msun$
(solid line).

{\bf $\star$ [Ni/Fe]}: 
\cite{fre08} reported an abundance ratio of Ni ([Ni/Fe] $=0.2$), that is
higher than HE~0107--5240. This suggests that a
larger amount of matter with low $\ye$ is ejected in the parent
SN of HE~1327--2326 than that of HE~0107--5240. Thus, the SN model of
HE~1327--2326 has deeper $\MCini$ than that of HE~0107--5240. 

{\bf $\star$ [Ti/Fe]}: 
Contrary to HE~0107--5240, Co in HE~1327--2326 is not detected and thus
the explosion energy of the parent SN of HE~1327--2326 is not
well constrained. Although the high [Ti/Fe] ratio is suggestive of 
high entropy, the enhancement of [Ti/Fe] is not enough even with $\E=5$. We
note that [\ion{Ti}{2}/\ion{Fe}{1}] ratio could have an ambiguity due to
a rather large NLTE effect \citep{ber11}. If [Ti/Fe] is ignored, the
abundance pattern of HE~1327--2326 can be explained by a low-$E$ SN model with
$\E=0.72$, $\Mbh=5.72\Msun$, and $\Mfe=1.53\times10^{-5}\Msun$ 
(dashed line).

\begin{figure*}
\epsscale{.9}
\plotone{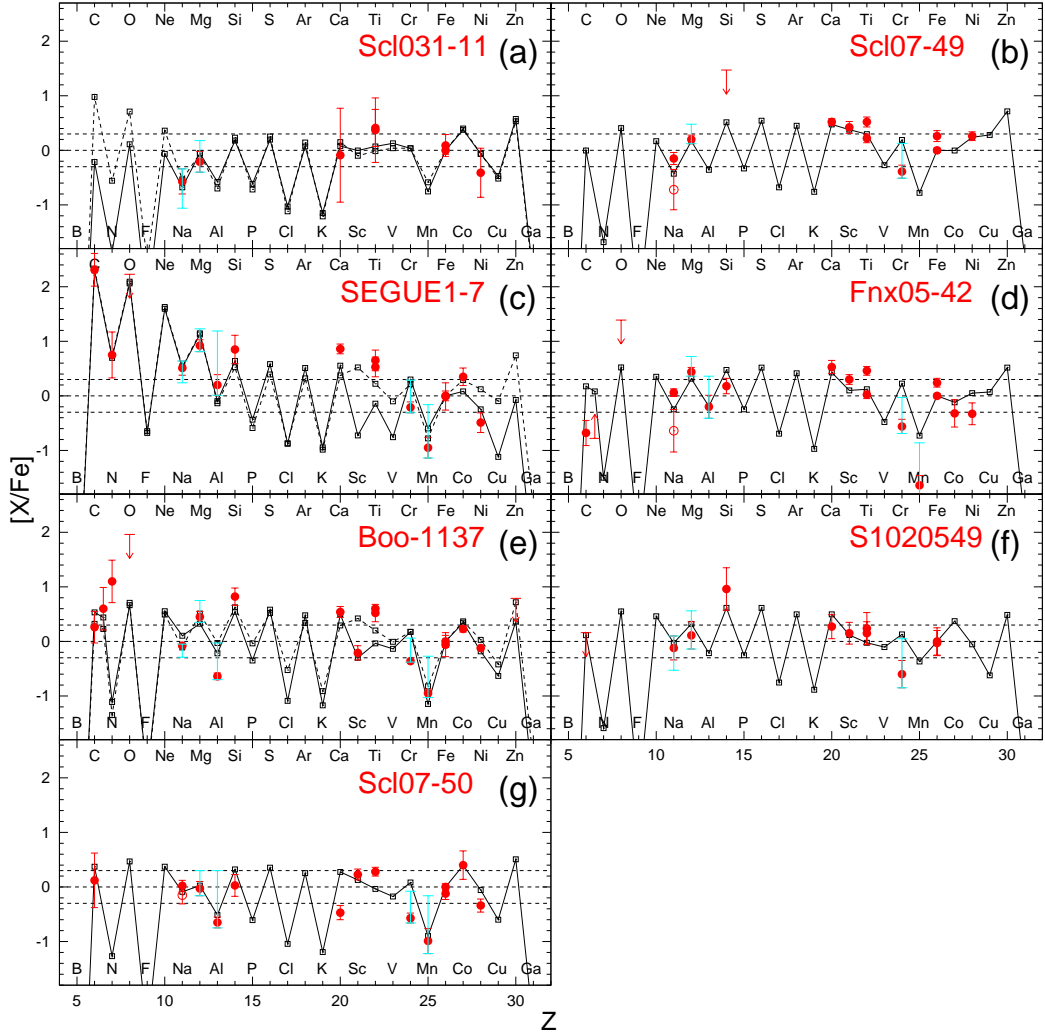}
\figcaption{Same as Fig.~\ref{fig:EMP1} but for the EMP stars in dwarf galaxies.\label{fig:dSph}}
\end{figure*}

\subsection{EMP stars in dwarf galaxies}
\label{sec:dsph}

EMP stars with [Fe/H]~$\lsim-3.5$ have recently been discovered in dwarf
galaxies and their detailed abundance ratios have become available. The
abundance patterns of EMP stars in dwarf galaxies are similar to those of
halo stars at low metallicity \citep[\eg][]{fre10}, in contrast to the
majority of stars
in dwarf galaxies at [Fe/H]~$\gsim-3$. Thus it is suggested that the
early chemical enrichment of all
galaxies may be nearly identical. Therefore, although the ISM density in
dwarf galaxies at the formation of the EMP stars could be different, it is
likely that the abundance pattern of the EMP star reflects dominantly nucleosynthesis in a
single SN even in dwarf galaxies. Although, owing to the observational
difficulty, the number of detectable lines is small and several
abundance ratios of the EMP stars have large errors, we first
construct Pop III SN models for the EMP stars in dwarf galaxies. 

{\bf Scl~031--11}: The metallicity is [Fe/H]~$=-3.47$ and the
abundance ratios are shown in Figure~\ref{fig:dSph}a \citep{sta12}. The
abundance ratios [(C,~N,~O)/Fe] are not available and error bars of
[(Ca,~Ti)/Fe] are large. The abundance pattern is well reproduced by an SN
model with $\E=20$, $\Mbh=3.26\Msun$, and $\Mfe=2.82\times10^{-1}\Msun$
(solid line). 

{\bf $\star$ [Mg/Fe]}: 
The low [Mg/Fe] can be explained without invoking the contribution of
Type Ia supernovae but only with a hypernova-like core-collapse
explosion. In the above model (solid line), the low [Mg/Fe] is resulted
from a large amount of Fe ejection which stems from large $E$ and
relatively small $\Mbh$. However, the constraints on $\Mfe$ and $\Mbh$
is not strict owing to the small number of measured abundance ratios and
the large errors. In fact, the abundance pattern could be explained by an SN model with
$\E=10$, $\Mbh=5.82\Msun$, and $\Mfe=1.34\times10^{-2}\Msun$ as well
(dashed line). In this model, much larger amount of fallback, \ie large
$\Mmout$ and $\fej$, causes smaller amount of ejected Mg, which leads to
the lower [Mg/Fe].

{\bf Scl~07--49}: The metallicity is [Fe/H]~$=-3.48$ and the
abundance ratios are shown in Figure~\ref{fig:dSph}b \citep{taf10}. The
abundance ratios [(C,~N,~O)/Fe] are not available. The
abundance pattern is well reproduced by an SN model with $\E=20$,
$\Mbh=2.63\Msun$, and $\Mfe=1.77\times10^{-1}\Msun$.

{\bf $\star$ [Mg/Fe]}: 
The low [Mg/Fe] is explained by large $\Mfe$.
However, as far as $E$ is high enough ($\E\gsim10$) to realize high
[(Sc,~Ti)/Fe], $\Mfe$ is not well
constrained as in the case of Scl~031--11.

{\bf SEGUE1-7}: The metallicity is [Fe/H]~$=-3.57$ and the
abundance ratios are shown in Figure~\ref{fig:dSph}c
\citep{nor10}. We adopt the LTE Mn abundance without an increase of 
$0.4$~dex. The abundance pattern except for [Ti/Fe] is well
reproduced by an SN model with $\E=20$, $\Mbh=5.72\Msun$,
and $\Mfe=7.55\times10^{-4}\Msun$ (solid line). Since SEGUE1-7 is a CEMP star, the SN model with
small $\Mfe$ reproduces the abundance pattern as in \S~\ref{sec:CEMP}. 

{\bf $\star$ [(Ca,~Ti,~Ni)/Fe]}: 
The high [Ca/Fe] and low [Ni/Fe] implies a large $\MCini$ which leads to
the underproduction of [Ti/Fe]. If we ignore [Ni/Fe] determined by a
single \ion{Ni}{1} $\lambda3858$~\AA\ line, the agreements of [Ti/Fe] and
[Co/Fe] are improved with an SN
model with $\E=20$, $\Mbh=5.72\Msun$, and $\Mfe=6.99\times10^{-4}\Msun$
(dashed line).

{\bf Fnx~05--42}: The metallicity is [Fe/H]~$=-3.66$ and the
abundance ratios are shown in Figure~\ref{fig:dSph}d
\citep{taf10}. The abundance pattern is well
reproduced by an SN model with $\E=20$, $\Mbh=3.36\Msun$,
and $\Mfe=1.18\times10^{-1}\Msun$.
The low [(Co,~Ni)/Fe] are
reproduced by a large $\MCini$.

{\bf $\star$ [C/Fe]}: 
Fnx~05--42 shows low [C/Fe]. The low
[C/Mg] is difficult to be reproduced by the SN model even with 
no fallback. However, since
there are no constraint on the abundance of N and Fnx~05--42 has low surface
gravity, C could be converted to N in Fnx~05--42. We thus adopt a lower
limit of [(C+N)/Fe] instead of [C/Fe], which is consistent with the SN model. 

{\bf Boo--1137}: The metallicity is [Fe/H]~$=-3.71$ and the
abundance ratios are shown in Figure~\ref{fig:dSph}e \citep{nor10boo}. 
Boo--1137 is an NEMP giant star and thus we adopt [(C+N)/Fe] instead of [C/Fe]
and [N/Fe]. We adopt the LTE Mn abundance without an increase of
$0.4$~dex. The abundance pattern except for [Ti/Fe] is well reproduced
by an SN model with $\E=5$,
$\Mbh=4.65\Msun$, and $\Mfe=4.49\times10^{-2}\Msun$ (solid line). 

{\bf $\star$ [Ti/Fe]}: 
The
underproduction of [Ti/Fe] stems from the low [Sc/Fe]. The disagreement
of [Ti/Fe] is improved with an SN model with higher $E$, \ie $\E=20$, $\Mbh=3.30\Msun$,
and $\Mfe=8.29\times10^{-2}\Msun$ if we
ignore [Sc/Fe] determined by three lines (dashed line). The inability to
simultaneously reproduce [Sc/Fe] and [Ti/Fe] might result from the NLTE
effect of [Ti/Fe] determined by \ion{Ti}{1} lines because [Sc/Fe] and
[Ti/Fe] in Boo--1137 is incongruous with the correlation between [Sc/Fe]
and [\ion{Ti}{2}/Fe] \citep{yon13}.

{\bf S~1020549}: The metallicity is [Fe/H]~$=-3.81$ and the
abundance ratios are shown in Figure~\ref{fig:dSph}f \citep{fre10}. 
For the CNO elements, only a low upper limit to 
[C/Fe] is obtained. The abundance pattern is well reproduced by an
SN model
with $\E=10$, $\Mbh=2.39\Msun$, and $\Mfe=1.43\times10^{-1}\Msun$. 
The low upper limit to [C/Fe] and
low [Mg/Fe] are explained by large $\Mfe$.

{\bf Scl~07--50}: The metallicity is [Fe/H]~$=-3.96$ and the
abundance ratios are shown in Figure~\ref{fig:dSph}g \citep{taf10}. The
abundance pattern except for [Ca/Fe] is well reproduced by an
SN model with $\E=10$,
$\Mbh=4.75\Msun$, and $\Mfe=6.96\times10^{-2}\Msun$. 

{\bf $\star$ [Ca/Fe]}: 
The observed
abundance of Ca is suspiciously low due to the \ion{Ca}{1}
$\lambda4227$~\AA\ resonance line sensitive to an NLTE effect
\citep{mas07,taf10}.

\begin{figure*}
\epsscale{.9}
\plotone{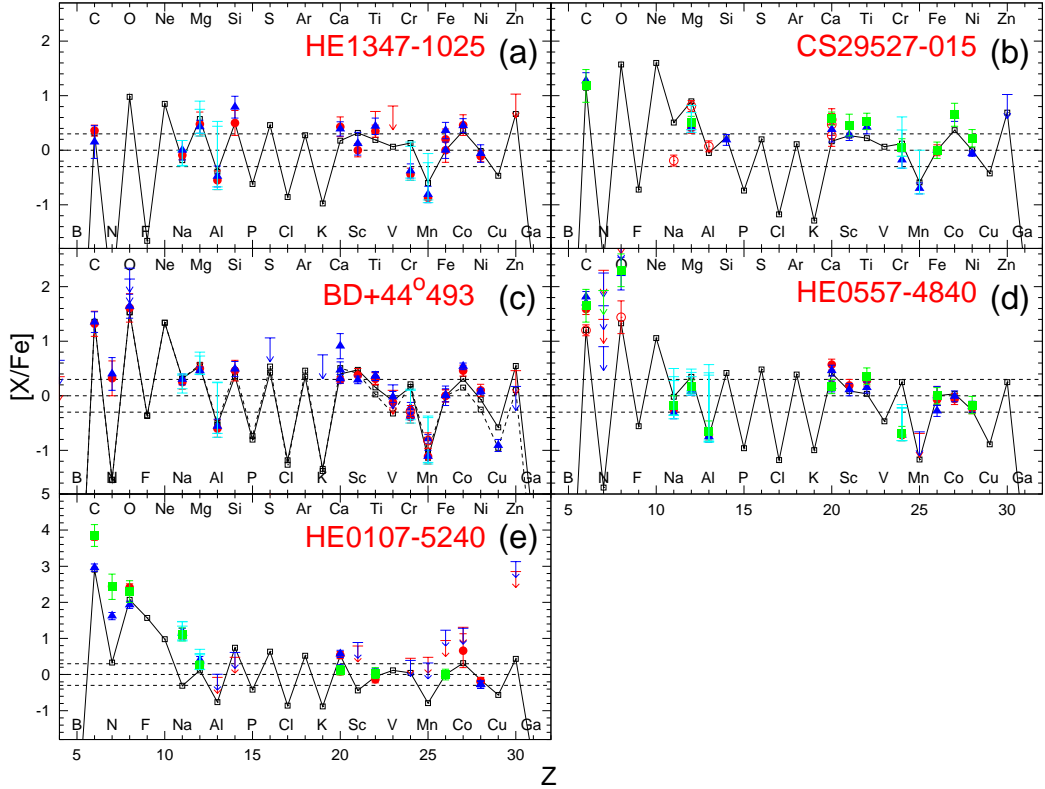}
\figcaption{Same as Fig.~\ref{fig:EMP1} but with the $40\Msun$
 SN models. The properties and legends of Pop III SN models are summarized
 in Table~\ref{tab:40Msun}. \label{fig:40Msun}}
\end{figure*}

\section{Limitation of abundance profiling}
\label{sec:limit}

\subsection{Progenitor mass}
\label{sec:40Msun}

Although we expediently construct the SN models with the $25\Msun$ progenitors, Pop
III SNe with various progenitor masses should contribute to the chemical
enrichment in the early Universe. In this subsection, we construct SN models for
HE~1347--1025 (the EMP star), CS~29527--015 and BD+44$^\circ$493 (the
CEMP stars), HE~0557--4840 (the UMP stars), and HE~0107--5240 (the HMP
stars) using the $40\Msun$ progenitor without enhancement of mixing (Model C).

The $40\Msun$ SN models are also in reasonable agreement with the
abundance patterns of the MP stars (Figs.~\ref{fig:40Msun}a-\ref{fig:40Msun}e). 
The model parameters are summarized in
Table~\ref{tab:40Msun}. Improvements made in the $25\Msun$ SN models are also possible for the
$40\Msun$ SN models as follows: 
\begin{itemize}
 \item The underproduction of N for BD+44$^\circ$493 and
HE~0107--5240 can be resolved if $10\%$ and $2\%$ of C in the He layer is
converted to N, respectively. 
 \item The overproduction of Cu and Zn for BD+44$^\circ$493 can be
improved by adopting larger $\MCini$, although it would make the agreement
of [(Ti,~Co,~Ni)/Fe] worse (dashed line). 
\end{itemize}

The differences between the $25\Msun$ and $40\Msun$ SN models, which can be
used to distinguish the progenitor masses, appear in (1)
[C/O] for HE~1347--1025, (2) [(Na,~Al)/Mg] for HE~1347--1025,
CS~29527--015, BD+44$^\circ$493, and HE~0557--4840, and (3) [Mg/Fe] for
HE~0107--5240. Compared with the $25\Msun$ models, the abundances of the
$40\Msun$ SN model show the following features.

{\bf (1) Low [C/O]:} This stems from the fact that the $40\Msun$
progenitor has a thicker O+Mg layer compared to a C+O layer than the
$25\Msun$ progenitor. The low [C/O] is prominent in the SN model with
small fallback, i.e., for the EMP stars. On the other hand, [C/O] in the
models for stars with high [C/Fe] is similar to the $25\Msun$ SN model
because the O mass is reduced by the large fallback of the O+Mg
layer. If the O abundance in HE~1347--1025 would be measured within
the error of 0.2~dex, we could distinguish between the $25\Msun$ and
$40\Msun$ progenitors.

{\bf (2) Low [(Na,~Al)/Mg]:} The $40\Msun$ progenitor has lower
Na/Mg and Al/Mg in the O+Mg layer due to C burning at higher
temperature than the $25\Msun$ progenitor. Therefore, the $40\Msun$ SN
models better-reproduce low
[Na/Mg] of HE~1347--1025, BD+44$^\circ$493, and HE~0557--4840
(Figs.~\ref{fig:40Msun}a, \ref{fig:40Msun}c, \ref{fig:40Msun}d) than the
$25\Msun$ SN models. The $40\Msun$ SN models also predict low [Al/Mg],
which is consistent with the LTE abundance ratio but lower than the NLTE-corrected ratio.
If the NLTE effect for [Al/Fe] is small, the $40\Msun$ SN models are
more favored for HE~1347--1025, BD+44$^\circ$493, and HE~0557--4840 than
the $25\Msun$ SN models.

For CS~29527--015, [Na/Mg] of the $40\Msun$ SN model is not low enough to
reproduce the observed ratio (Fig.~\ref{fig:40Msun}b) and [Na/Mg] of the
$25\Msun$ SN model is clearly too high (Fig.~\ref{fig:CEMP}e). For
[Al/Mg] of CS~29527--015, on the other hand, both the $40\Msun$ and
$25\Msun$ SN models are consistent with the observation. Such
consistency in [Al/Mg] between both models and CS~29527--015 stems from
the enhancement of ejection of explosively-synthesized Al in the
$40\Msun$ SN model. This results from the
high explosion energy and relatively small $\Mmout$ that 
locates at the middle of O+Mg layer.

{\bf (3) Low [Mg/Fe]:} The $40\Msun$ SN model for HE~0107--5240 shows lower
[Mg/Fe] ($\sim0$) than the $25\Msun$ SN model (Figs.~\ref{fig:HMP}a and
\ref{fig:40Msun}e). Extended fallback reaching to
the outer edge of the O+Mg layer, \ie large $\Mmout$, is required to
high [C/Mg], while relatively small $\MCini$ is required to reproduce
low [Si/Fe] and high [Co/Fe]. These requirements lead to lower [Mg/Fe] of
the $40\Msun$ SN model than the observation. Thus, the $25\Msun$ SN model is favored for
HE~0107--5240.

The above examples demonstrate that it is possible to constrain the
progenitor mass by using the abundance ratios [C/O], [(Na,~Al)/Mg],
and [Mg/Fe]. However, there are still theoretical and
observational ambiguities. For example: (1) Mixing efficiency in the stellar
evolution would affect the abundance ratios among the CNO elements. (2)
The 3D and NLTE effects would affect the abundance
determination. We emphasize that, in order to constrain the progenitor
mass from the abundance pattern of each MP star, it is important to
obtain the elemental abundances including 3D-NLTE effects as many as
possible.

\subsection{Explosion energy}
\label{sec:energy}

In spherical explosion models, the temperature and entropy during explosive
nucleosynthesis and the amount of fallback are determined
predominantly by the explosion energy
for a given progenitor model. This is one of the reason why the explosion
energy is treated as an important property to rule explosive
nucleosynthesis in previous studies. However, in multi-dimensional
explosions, the temperature, entropy, and the amount of fallback can be
different even for the same explosion energy or could be similar even
for the different explosion energies. For example, the resultant SN
yields can be different even with the same explosion energies \citep{tom07a}.

Figures~\ref{fig:EMP1}d and \ref{fig:EMP2}d show that the SN models with $\E=5$ and $20$
explain the abundance patterns of CS~22897--008 with the progenitor Model B
and BS16467-062 with Model A (see also SDSS~J102915+172927 in Figure~\ref{fig:UMP}b
for the case with the small number of abundance determination). The
abundance patterns of these SN models are similar, \ie the
$E$-dependence is small, except for [F/Fe] in the progenitor Model B.
As discussed for CS~29498--043 in \S~\ref{sec:CEMP},
F is enhanced via the hot CNO cycle
$^{14}$N($p$,$\gamma$)$^{15}$O($\alpha$,$\gamma$)$^{19}$Ne($\beta^+$)$^{19}$F
during explosive nucleosynthesis if the N-rich layer experiences 
$T\gsim7\times10^8$~K (\eg \citealt{wie10}).\footnote{The required
temperature is higher than the value in \cite{wie10} due to the short
cooling time scale in explosive nucleosynthesis. Explosive
nucleosynthesis even at $T\lsim10^9$~K is important for synthesis of
$^{11}$B and $^{19}$F in the He/N-rich layer of the H-envelope.}
Although NIR spectroscopy is required to determine the abundance of F
(\eg \citealt{smi05}), [F/Fe] could be an additional clue to constrain
the explosion energy.

\begin{figure*}
\epsscale{.7}
\plotone{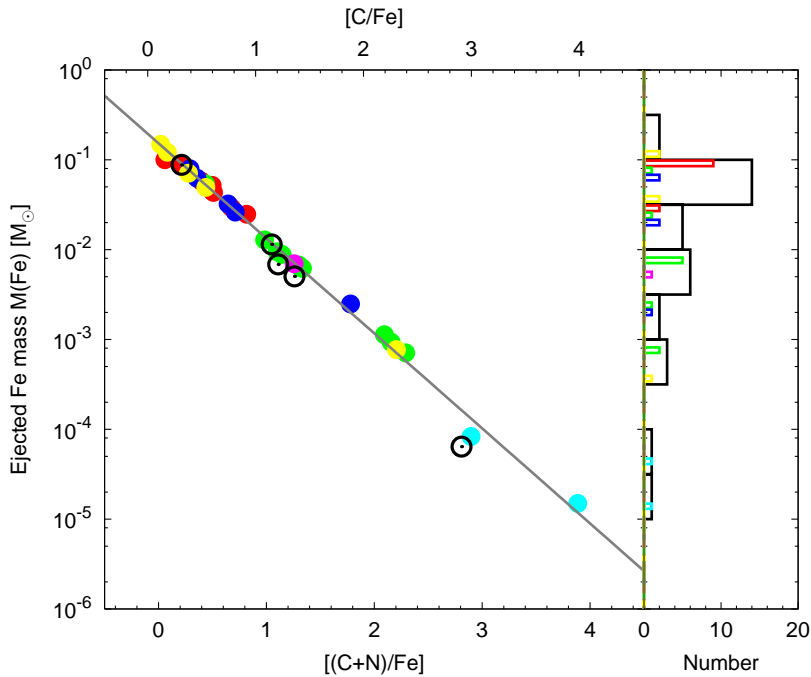}
\figcaption{(Left) Ejected Fe masses of SN models for the individual
 MP stars as a function of [(C+N)/Fe]. The color represents the
 EMP stars ({\it red}), the CEMP stars ({\it green}), the NEMP stars ({\it
 blue}), the UMP stars ({\it magenta}), the HMP stars ({\it cyan}), and
 the EMP stars in dwarf galaxies ({\it yellow}). The fitting curve
 [Eq.~(\ref{eq:mfe})] is also shown
 in {\it gray line}. The black open circles
 represent the $40\Msun$ SN models shifted by a factor of
 $\Mc_{25\Msun}/\Mc_{40\Msun}$ according to
 Equation~(\ref{eq:mfe}). 
 (Right) Number distribution of ejected Fe masses of Pop III SNe. The colors
 represent the same as in the left panel but the black includes
 $25\Msun$ SN models for all kinds of MP stars. \label{fig:cfe}}
\end{figure*}

The weak constraint on the explosion energy stems from the fact that we
allow arbitrary choices of the mixing-and-fallback parameters and the low-density
modification. This is
an inevitable shortcoming in the 1D model that approximates the aspherical explosion
with the mixing-and-fallback
model. In reality, the temperature, entropy, and fallback are determined by
energy injection, geometry, \eg a jet-opening angle, and fraction of
kinetic energy in the jet-induced explosion
\citep[\eg][]{tom09a}. Therefore, we leave the constraint on the
explosion energy for future multi-dimensional studies. 

\vspace{1cm}

\section{Distribution of supernova properties}
\label{sec:dist}

According to the Pop III SN models constructed in the previous sections, we
investigate how properties of Pop III SNe link to the abundance ratios. 
In order to reduce the uncertainties due to the lack of observations, we focus
on the 34 Pop III SN models reproducing the EMP stars with determined
abundance ratios or strict upper limit of 4 elemental groups
listed in \S~\ref{sec:obs}, except for
HE~0057--5959.\footnote{HE~0057--5959 could be originated from the mass
transfer from the N-rich AGB companion.} The adopted models are indicated by ``Yes''
in Table~\ref{tab:models}. We employ theoretical abundance ratios
instead of observed abundance ratios to reduce ambiguities due to
3D-NLTE effects. 

The following subsections proceed as follows.

First, we derive the relation between the properties of
observationally-selected theoretical models and the abundance ratios of
their yields. The relation itself is free from the observational
uncertainties such as 3D-NLTE effects, although the requirements to the
SN models from the EMP stars could be changed by 3D-NLTE effects.

Second, we derive the distribution of properties of Pop III
SNe. The observational uncertainties and poor reproduction by
theoretical models influence the distribution of properties. 
In addition to these limitations, the distribution
also involves the following assumptions; (1) the detection of EMP stars at
[Fe/H]~$\lsim-3.5$ is observationally unbiased and (2) all Pop III SNe
contribute to the EMP stars at [Fe/H]~$\lsim-3.5$ with the same
probability. We discuss validity of these assumption in
\S~\ref{sec:valid}.

\begin{figure*}
\epsscale{.7}
\plotone{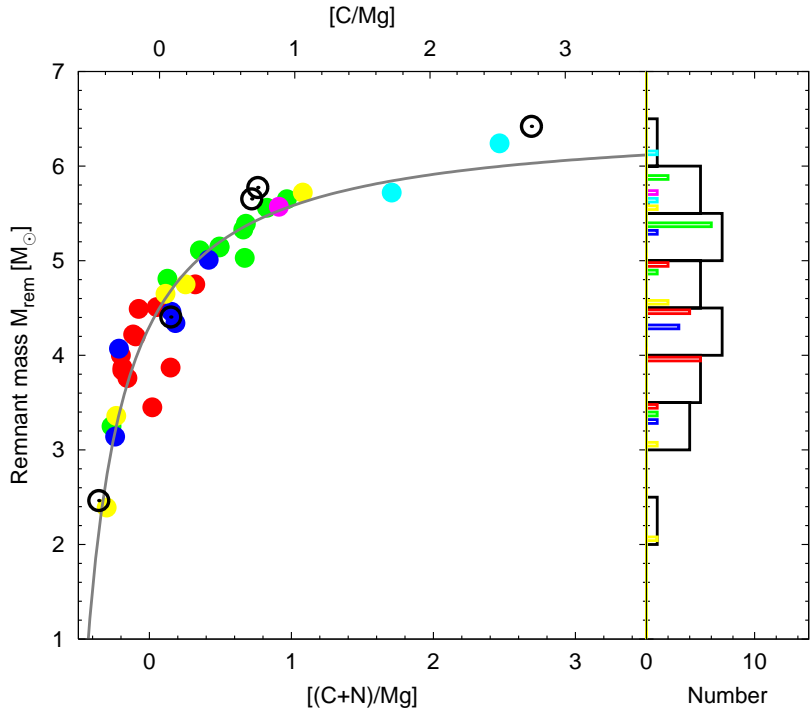}
\figcaption{(Left) Remnant masses of SN models for the individual
 MP stars as a function of [(C+N)/Mg]. The color represents the
 same as in the left panel of Figure~\ref{fig:cfe}. The fitting curve
 [Eq.~(\ref{eq:mrem})] is also shown
 in {\it gray line}. The black open circles
 represent the $40\Msun$ models shifted by a factor of
 $M_{{\rm C+O},25\Msun}/M_{{\rm C+O},40\Msun}$ according to
 Equation~(\ref{eq:mrem}). 
 (Right) Number distribution of remnant masses of Pop III SNe. The colors
 represent the same as in the right panel of Figure~\ref{fig:cfe}. \label{fig:cmg}}
\end{figure*}

\vspace{1cm}
\subsection{[(C+N)/Fe] vs. $\Mfe$}
\label{sec:cfe}

The ejected Fe mass $\Mfe$ is an important property of the SN because
energy release from the radioactive decay of parent \Nifs\ powers the
light curve of the SN. [(C+N)/Fe] of MP stars constrain the ratio
between $\Mfe$ and the ejected mass of C+N, $\Mec$, of Pop III SNe. Since 
some fractions of C and N are destroyed by explosive nucleosynthesis and fall back onto
the central remnant, $\Mec$ is reduced from the mass of C+N in the SN
progenitor [$\Mc$] and such a reduction occurs more efficiently for the
SN models with larger amount of fallback.

The left panel of Figure~\ref{fig:cfe} demonstrates that [(C+N)/Fe]
relates to $\Mfe$. The relation is well fitted by the following equation
\begin{eqnarray}
\nonumber \Mfe =0.931 \left({X_{\rm Fe}\over{X_{\rm C}+X_{\rm
 N}}}\right)_\odot \\
 \times \Mc \times 10^{-1.058{\rm [(C+N)/Fe]}}, \label{eq:mfe}
\end{eqnarray}
where $X_{\rm Fe}$, $X_{\rm C}$, and $X_{\rm N}$ are the mass fractions of
Fe, C, and N,\footnote{
$\left({X_{\rm Fe}\over{X_{\rm C}+X_{\rm N}}}\right)_\odot=0.423$ \citep{asp09}.}
respectively.
The ratio $\Mec/\Mc$ ranges over $0.545-0.931$ for [(C+N)/Fe]$=-4$
to $0$. We note that
[C/Fe]$\simeq$[(C+N)/Fe]$+0.1$ for EMP stars without N enhancement \citep{asp09}.

$\Mc$ depends on $\Mms$ of the SN progenitor. For example,
among the models in \cite{ume05a} (see also \citealt{tom07b}),
$\Mc=0.07\Msun$ and $1.8\Msun$ for the $13\Msun$ and $50\Msun$
progenitors, respectively. The $40\Msun$ models constructed in
\S~\ref{sec:40Msun} are also shown in the left panel of Figure~\ref{fig:cfe}, shifting
by a factor of $\Mc_{25\Msun}/\Mc_{40\Msun}$ ($=0.592$) according to
Equation~(\ref{eq:mfe}), where $\Mc_{25\Msun}$ ($=0.388\Msun$) and
$\Mc_{40\Msun}$ ($=0.656\Msun$) are
the mass of C+N in the $25\Msun$ and $40\Msun$ progenitors,
respectively. The $40\Msun$ SN models are consistent
with the $25\Msun$ SN models. The variation of $\Mc$ in the progenitor
models with $\Mms=13-50\Msun$ ($\sim1.5$~dex)
is smaller than that of [(C+N)/Fe] in the EMP stars ($\sim4$~dex). Therefore, the real
variation of $\Mfe$ amounts at least to $\sim2.5$~dex, which would reflect
the differences in the explosion geometry, explosion energy, and fallback.

We find from Figure~\ref{fig:cfe} (right) that the most common Pop III SN yields
$\Mfe=3\times10^{-2}-10^{-1}\Msun$, which 
corresponds to the peak bolometric
magnitude of Type Ibc SNe $\Mpeak$ of $-16.6$ to
$-17.4$~mag.\footnote{The peak brightness of a Type Ibc SN is
roughly proportional to $\Mfe$. A Type Ibc SN with $\Mfe=0.07\Msun$ has
$\Mpeak$ of $-17$~mag \citep[\eg SN~2008D,][]{tanaka09} independent of
the metallicity.} This is consistent with the brightness of the most
common Type Ibc SN in the present day (\eg $\Mpeak\sim-17.5$~mag,
\citealt{ric02}; see also \citealt{li11part2} in which 
host galaxy extinction is not corrected).

On the other hand, the number distribution of $\Mfe$ in the right panel of
Figure~\ref{fig:cfe} indicates an extended tail for 
Pop III SNe down to $\Mfe\sim10^{-2}-10^{-5}\Msun$.\footnote{The
tail would be more extended if the fallback reaches the He layer
that contains C and N.} Such a tail
corresponds to Type Ibc SNe
with $\Mpeak>-14.9$~mag, which is 2~mag fainter than the most common
Type Ibc SNe in the present day. Such faint Type Ibc SNe (\eg 
SN~2008ha, \citealt{val09,fol10}) and GRBs without associated SNe \citep[\eg
GRB~060614,][]{fyn06} have been discovered in the nearby
Universe. However, the fraction of such faint Type Ibc SNe is very small
in the present day, in contrast to a non-negligible fraction of
Pop III SNe with $\Mfe\lsim10^{-2}\Msun$. 
Such a discrepancy in the fractions could stem from the following biases;
(1) There exists an observational bias in the nearby Universe due to the
dimness of faint Type Ibc SNe. The fraction of dim SNe in the nearby
Universe is an interesting issue to be addressed \citep[\eg][]{hor11}.
(2) There may be a bias that the EMP stars are enriched from SNe of
massive stars while nearby SNe are mainly the explosions of less-massive
stars. 

Future nearby SN surveys and observations will solve these biases with
deep observations of nearby SNe and the increase of the number of nearby
SNe of massive stars. If the discrepancy in fraction is real, it 
implies that the Pop III SNe tend to have smaller $\Mfe$ than the
present day SNe. This could be explained by the larger fallback in Pop
III SNe which is led by an existence of thicker H envelope due to the
weaker mass loss and a more compact structure due to the lower opacity
in a Pop III star.

\subsection{[(C+N)/Mg] vs. $\Mrem$}
\label{sec:cmg}

The remnant mass $\Mrem$ of the SN model is determined by
Equation~(\ref{eq:mfmodel}). $\Mrem$ is larger for higher [(C+N)/Mg]
because high [(C+N)/Mg] requires large $\Mmout$ (being close to the outer
boundary of the C+O core) and small $\fej$. The relation between $\Mrem$ and
[(C+N)/Mg] is shown in the left panel of Figure~\ref{fig:cmg}. The relation is well fitted
by the following formula
\begin{equation}
 \Mrem = \Mco \left(1-{0.241\over{{\rm [(C+N)/Mg]}+0.716}}\right), \label{eq:mrem}
\end{equation}
where $\Mco$ is the C+O core mass.
The $\Mco$ (and thus resulting $\Mrem$) depends on the progenitor mass,
\eg $M_{{\rm C+O},13\Msun}=2.4\Msun$, $M_{{\rm C+O},25\Msun}=6.49\Msun$, 
$M_{{\rm C+O},40\Msun}=14.0\Msun$, and $M_{{\rm C+O},50\Msun}=18.3\Msun$,
for the progenitors of $13\Msun$, $25\Msun$, $40\Msun$, and $50\Msun$,
respectively. The results for the
$40\Msun$ SN models constructed in
\S~\ref{sec:40Msun} are also shown in the left panel of Figure~\ref{fig:cmg}, shifting by
a factor of $M_{{\rm C+O},25\Msun}/M_{{\rm C+O},40\Msun}$ ($=0.462$) according to
Equation~(\ref{eq:mrem}). The relation for the $40\Msun$ SN models is consistent
with that for the $25\Msun$ SN models. The parent Pop III SNe of the EMP
stars with high [C/Mg] ($\gsim+3$) are
black-hole-forming\footnote{It depends on the maximum mass of
a neutron star \citep[\eg][]{dem10}.} SNe with $\Mrem\sim2.4\Msun$ for the $13\Msun$ star
and $\sim18.3\Msun$ for the $50\Msun$ star.

According to the $25\Msun$ SN models, $\Mrem$ widely distributes from
$3\Msun$ to $6\Msun$.\footnote{The maximum $\Mrem$ would be larger if
the materials in the He layer fall back.} Although it is difficult to
constrain $\Mrem$ from the observations of nearby SNe because of
the uncertainty in the fallback mass, the mass distribution of
stellar-mass black hole is constrained from the observations of X-ray binary
systems. The width of $\Mrem$ distribution is similar to that of the
present day mass
distribution of stellar-mass black holes \citep{oze10,far11}. We note
that the distribution of $\Mrem$ is extended 
depending on the initial mass function of Pop III stars. Further
investigation including an integration over the initial mass function is
required to quantitatively evaluate the consistency between the
distribution of $\Mrem$ of Pop III SNe and present day stellar-mass
black hole masses.

\subsection{Validity of assumptions}
\label{sec:valid}

The above methods stand on the assumptions: (1) the detection of
EMP stars at [Fe/H]~$\lsim-3.5$ is unbiased toward any subclasses
categorized in this paper and (2) all Pop
III SNe contribute to any subclasses of EMP stars with [Fe/H]~$\lsim-3.5$
at the same possibility. 

The assumption (1) is appropriate because the MP star surveys are performed with Ca
lines not with \eg CH or NH line, and follow-up spectroscopic 
observations are
complete at [Fe/H]~$\lsim-3.5$. The assumption (2) could be of no
matter if the mass of gas enriched by Pop III SNe and the initial mass
function of the next-generation stars do not depend on the metallicity at
[Fe/H]$\lsim-3.5$. The 3D computations propose wide distribution of
[Fe/H] from $-5$ to $-1$ but the gas mass seems to slowly increase as a
function of metallicity \citep{rit12}.\footnote{The 3D results 
depend on the details of simulations, and it is required to confirm the
dependence of gas metallicity on properties, \eg explosion energy, of
Pop III SNe with systematic works.} The dependence of initial mass function on metallicity is
under investigation \citep[\eg][]{omu05,dop12}. Furthermore the
fragmentation mechanisms of clouds are still under
debate (fine-structure cooling, \citealt{bro03} and dust-induced cooling,
\citealt{sch06}) and it is recently suggested that low-mass stars can
form even at $Z=0$ by disk fragmentation \citep{cla11,sus13}. Convolving the metallicity
distribution of gas mass and the initial mass function with Equations
(\ref{eq:mfe}) and (\ref{eq:mrem})
gives realistic distribution of properties of Pop III SNe through the
comparison with the distribution of the abundance ratios of the EMP stars.

\section{Discussion}
\label{sec:discussion}

The SN models well reproduce the abundance patterns of most of EMP
stars. However, there are several difficult elements. 
\begin{description}
\item[(1)] The high
[(Sc,~Ti)/Fe] of the EMP stars require the high-entropy environment in
the explosion. The variation of $\frho$ implies that Pop III SNe are
aspherical explosions with different energy deposition rate and/or
jet-opening angle \citep{tom09a}.
\item[(2)] Although the $\ye$ profile is
modified referring only [(Mn,~Co)/Fe] in this paper, the abundance
ratios of elements synthesized in the complete Si-burning layer could
minutely constrain the $\ye$ distribution in the innermost layer of the
SN ejecta. 
\item[(3)] The ratio [K/Fe] of the present SN models are not in
agreement with [K/Fe] of all the EMP stars. Although the NLTE effect on
[K/Fe] is large at [Fe/H]~$\sim-2$ to $-1$ \citep{tak02,zha06b}, the
effect at low metallicity is not large enough to reconcile the
discrepancy. There are several suggestions to enhance [K/Fe]. The
underproduction of K could be solved by nucleosynthesis in $p$-rich
ejecta which may be led by neutrino process \citep{iwa06}.
\end{description} 
We leave these issues in future investigation.

The determination of the abundance ratios in the EMP stars is
suffered from the 3D-NLTE effects and the difficult determination of the surface
gravity. An impact on the abundance ratios varies depending on individual stars.
The uncertainties of the abundance ratios make it difficult to minutely constrain
properties of SN models. For example, the uncertainties
for HE~1300+0157 are as large as $\sim2.0$~dex for [(C,~O)/Fe] and
$\sim1.0$~dex for [(Na,~Mg,~Al)/Fe]. Such large uncertainties lead large
ambiguities of $\Mfe$ by two orders of magnitude and $\Mrem$ by a factor
of $\sim2.5$. The low [$\alpha$/Fe] for the surface gravity of a dwarf
star can be reproduced even without the mixing-and-fallback process. 

We find that the progenitor mass can be constrained by the abundance ratios
between the elements synthesized in the same layer. In particular, the accurate
determination of the abundance ratios among Na, Mg, and Al could contribute
to the determination of the progenitor mass. We also indicate that the
abundance of F is useful for the constraint on $E$ if N is
enhanced in the progenitor and the N-rich layer experiences high
temperature ($T\gsim7\times10^8$~K) during the shock propagation. 
However, we caution that it is still difficult to uniquely determine the
progenitor mass and explosion energy with arbitrary choice of the
mixing-and-fallback
parameters because supernova nucleosynthesis also depends
on the explosion geometry. 2D or 3D explosion simulations
tie the explosion geometry, the explosion energy, and the main-sequence
mass, and help deriving physical insight on the central
engine. Furthermore, the abundance ratios in the progenitor
strongly depends on mixing processes in the SN progenitor, which are required to be dealt
with 3D computation \citep[\eg][]{mea07}. It is important to adopt SN
models with attention to these caveats. 

The number fraction of stars with low [(C+N)/Fe] in dwarf galaxies is larger than
that in Galactic halo and thus $\Mfe$ for the EMP stars in dwarf galaxies is distributed to 
higher values than those in halo stars. We note that the
smaller fraction of the EMP stars with high [(C+N)/Fe] in dwarf galaxies
is not explained by the observational bias. However, there are caveats that
the number of EMP stars with [Fe/H]~$\lsim-3.5$ in dwarf galaxies are
still small and that the abundance
of N is not well constrained except for two stars.
Although this might stem from statistical scatter, if
real, the Pop III SNe in dwarf galaxies yield low [$\alpha$/Fe] which
is consistent with the MP stars in dwarf galaxies at
[Fe/H]~$\gsim-3$. This would be verified by future studies with
increasing the number of EMP stars in dwarf galaxies.

\begin{figure}
\epsscale{1.}
\plotone{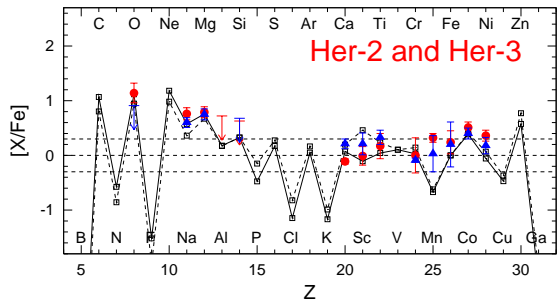}
\figcaption{Same as Fig.~\ref{fig:EMP1} but for the stars in Hercules dwarf
 galaxy. The properties and legends of Pop III SN models are summarized
 in Table~\ref{tab:models}. \label{fig:Her}}
\end{figure}

The chemical enrichment in dwarf galaxies is suggested to be
inhomogeneous even at higher metallicity [Fe/H]$\gsim-3$
\citep[\eg][]{koc08,ven12} because the abundance ratios of MP
stars in dwarf galaxies show large dispersion. The abundance patterns of stars formed in
the inhomogeneous environment can reflect predominantly nucleosynthesis in
a single SN even if the metallicity is high. Indeed, the
unusual abundance patterns of Her-2 and Her-3 stars with [Fe/H]~$=-2$
are reproduced by Pop III SN models with high explosion energies
(Fig.~\ref{fig:Her}). If the inhomogeneity of ISM is held at
high metallicity, a larger number of stars can be applied to constrain
the properties of Pop III SNe in the dwarf galaxies. It is important to
find out the transition period from inhomogeneous
to homogeneous metal enrichment in dwarf galaxies.

\section{Concluding Remarks}
\label{sec:conclusion}

We construct the Pop III SN models reproducing the abundance patterns of
the 48 EMP stars with [Fe/H]~$\lsim-3.5$ in the Galactic halo and dwarf
galaxies ({\sl abundance profiling}). We adopt the spherical explosion
models in which the aspherical
effects are taken into account as {\sl mixing-and-fallback}. According
to the SN models, we reveal the correlations between the
abundance ratios and the properties of Pop III SNe and derive the fitting formulae of the
relations: [(C+N)/Fe] vs. $\Mfe$, and [(C+N)/Mg] vs. $\Mrem$. 
This is enabled by the
construction of Pop III SN models for numerous EMP stars which is
impossible to be executed with 2D or 3D simulations because of 
computational costs. 
With the use of these relations, the distribution of the abundance
ratios of the EMP stars is projected to the distribution of
$\Mfe$ and $\Mrem$ of Pop III SNe. These clarify that the distribution
of $\Mfe$ extends down to $\Mfe\sim10^{-5}\Msun$, which might imply that
the Pop III SNe tend to have smaller $\Mfe$ than the present day SNe,
and that the distribution
of $\Mrem$ is as wide as $3\Msun$ which is consistent with the mass
distribution of present day stellar-mass black holes.

These
demonstrate that the abundance profiling is a powerful tool to derive the
properties of Pop III SNe. The future surveys and follow-up observations
can increase the number of EMP stars at [Fe/H]~$\lsim-3.5$ and make it
possible to more rigorously constrain the properties of Pop III SNe.
We emphasize the importance of the precise
determination of the abundance ratios including the 3D-NLTE effects and
the stellar parameters. These are essential to constrain the properties of Pop
III SNe including $E$ and $\Mms$ from the abundance pattern of
individual EMP stars by abundance profiling.

\acknowledgments

The authors thank Chiaki Kobayashi, Hajime Susa, and Marco Limongi for fruitful discussion on
chemical abundance, star formation in the early Universe, and synthesis
of F, respectively.
Data analysis were in part carried out on the general-purpose PC farm
at Center for Computational Astrophysics, National Astronomical
Observatory of Japan.
This research has been supported in part by World Premier
International Research Center Initiative, MEXT,
Japan, and by the Grant-in-Aid for Scientific Research of the JSPS
(20540226, 21840055, 23224004, 23740157) and MEXT (19047004, 22012003).


\begin{deluxetable}{ccccc}
 \tabletypesize{\scriptsize}
 \tablecaption{The progenitor models
 \label{tab:preSN}}
 \tablewidth{0pt}
 \tablehead{
   \colhead{Model name}
 & \colhead{$\Mms$}
 & \colhead{Mixing enhancement}
 & \colhead{Figure}
 & \colhead{Reference}\\
   \colhead{}
 & \colhead{[$\Msun$]}
 & \colhead{}
 & \colhead{}
 & \colhead{}
 }
\startdata
A & 25 & No  & Fig.~\ref{fig:preSN}a & \cite{iwa05} \\
B & 25 & Yes & Fig.~\ref{fig:preSN}b & \cite{iwa05} \\
C & 40 & No  & Fig.~\ref{fig:preSN}c & \cite{tom07a}
\enddata

\end{deluxetable}

\begin{deluxetable}{ccl}
 \tabletypesize{\scriptsize}
 \tablecaption{EMP stars with [Fe/H]~$\lsim-3.5$
 \label{tab:stars}}
 \tablewidth{0pt}
 \tablehead{
   \colhead{Star name}
 & \colhead{[Fe/H]}
 & \colhead{Figure: ({\it color}) reference}
 }
\startdata
\multicolumn{3}{c}{EMP stars} \\ \hline
HE~0146--1548       & $-$3.46 & Fig.~\ref{fig:EMP1}a: ({\it red}) 1,2 \\
CS~22189--009       & $-$3.48 & Fig.~\ref{fig:EMP1}b: ({\it red}) 3,4,5,6,7,8,9, ({\it blue}) 1 \\
SDSS~J090733+024608 & $-$3.50 & Fig.~\ref{fig:EMP1}c: ({\it red}) 10 \\
CS~22897--008       & $-$3.50 & Fig.~\ref{fig:EMP1}d: ({\it red}) 3,4,5,6,7,8,9, ({\it blue}) 1 \\
BS~16545--089       & $-$3.50 & Fig.~\ref{fig:EMP1}e: ({\it red}) 11, ({\it blue}) 12, ({\it green}) 13, ({\it magenta}) 1 \\
CS~22963--004       & $-$3.54 & Fig.~\ref{fig:EMP1}f: ({\it red}) 13, ({\it blue}) 1 \\
CS~22968--014       & $-$3.58 & Fig.~\ref{fig:EMP1}g: ({\it red}) 3,4,5,6,7,8,9, ({\it blue}) 1 \\
HE~1356--0622       & $-$3.58 & Fig.~\ref{fig:EMP1}h: ({\it red}) 14, ({\it blue}) 1 \\
HE~1347--1025       & $-$3.62 & Fig.~\ref{fig:EMP1}i: ({\it red}) 14, ({\it blue}) 1 \\
BS~16076--006       & $-$3.66 & Fig.~\ref{fig:EMP1}j: ({\it red}) 15, ({\it blue}) 1 \\
CS~22876--032AB     & $-$3.66 & Fig.~\ref{fig:EMP2}a: ({\it red}) 16 (Star A), ({\it blue}) 16 (Star B) \\
HE~0049--3948       & $-$3.68 & Fig.~\ref{fig:EMP2}b: ({\it red}) 1 \\
HE~1320--2952       & $-$3.69 & Fig.~\ref{fig:EMP2}c: ({\it red}) 1 \\
BS~16467--062       & $-$3.74 & Fig.~\ref{fig:EMP2}d: ({\it red}) 3,4,5,6,7,8,9,
 ({\it blue}) 14, ({\it green}) 13, ({\it magenta}) 1 \\
CS~22172--002       & $-$3.74 & Fig.~\ref{fig:EMP2}e: ({\it red}) 3,4,5,9, ({\it blue})
 17, ({\it green}) 1\\
HE~0228--4047       & $-$3.75 & Fig.~\ref{fig:EMP2}f: ({\it red}) 1 (dwarf), ({\it blue}) 1 (subgiant) \\
G~238-30            & $-$3.77 & Fig.~\ref{fig:EMP2}g: ({\it red}) 18, ({\it blue}) 19 \\
CS~22885--096       & $-$3.77 & Fig.~\ref{fig:EMP2}h: ({\it red}) 3,4,5,6,7,8,9,
 ({\it blue}) 17, ({\it green}) 1 \\  \hline
\multicolumn{3}{c}{CEMP stars} \\ \hline
HE~1150--0428       & $-$3.47 & Fig.~\ref{fig:CEMP}a: ({\it red}) 20,
 ({\it blue}) 1 \\
HE~1012--1540       & $-$3.48 & Fig.~\ref{fig:CEMP}b: ({\it red}) 14,
 ({\it blue}) 1 \\
CS~29498--043       & $-$3.52 & Fig.~\ref{fig:CEMP}c: ({\it red}) 21,
 ({\it blue}) 1 \\
HE~1506--0113       & $-$3.54 & Fig.~\ref{fig:CEMP}d: ({\it red}) 1 \\
CS~29527-015        & $-$3.55 & Fig.~\ref{fig:CEMP}e: ({\it red}) 6,7,8,9,
 ({\it blue}) 1 \\
HE~0132--2429       & $-$3.70 & Fig.~\ref{fig:CEMP}f: ({\it red}) 14,
 ({\it blue}) 1 \\
HE~1300+0157        & $-$3.76 & Fig.~\ref{fig:CEMP}g: ({\it red}) 22 (subgiant),
 ({\it blue}) 22 (dwarf), ({\it green}) 23, ({\it magenta}) 14, ({\it yellow}) 1 \\
BD+44$^\circ$493    & $-$3.78 & Fig.~\ref{fig:CEMP}h: ({\it red}) 24,
 ({\it blue}) 25 \\
HE~1201--1512       & $-$3.89 & Fig.~\ref{fig:CEMP}i: ({\it red}) 1 (dwarf),
 ({\it blue}) 1 (subgiant) \\
HE~2139--5432       & $-$4.02 & Fig.~\ref{fig:CEMP}j: ({\it red}) 1 \\
53327-2044-515      & $-$4.04 & Fig.~\ref{fig:CEMP}k: ({\it red}) 1,2 (dwarf),
 ({\it blue}) 1,2 (subgiant) \\
G~77--61            & $-$4.08 & Fig.~\ref{fig:CEMP}l: ({\it red}) 26,27 \\ \hline
\multicolumn{3}{c}{NEMP stars} \\ \hline
CS~22952--015       & $-$3.44 & Fig.~\ref{fig:NEMP}a: ({\it red}) 3,4,5,6,7,8,9,
 ({\it blue}) 28, ({\it green}) 1 \\
BS~16550--087       & $-$3.54 & Fig.~\ref{fig:NEMP}b: ({\it red}) 13,
 ({\it blue}) 1 \\
CS~22949--037       & $-$3.89 & Fig.~\ref{fig:NEMP}c: ({\it red})
 3,4,5,6,7,8,9,29, ({\it blue}) 17,30, ({\it green}) 14, ({\it magenta}) 1 \\ 
CS~30336-049        & $-$4.07 & Fig.~\ref{fig:NEMP}d: ({\it red}) 13,
 ({\it blue}) 1 \\
HE~0057--5959       & $-$4.08 & Fig.~\ref{fig:NEMP}e: ({\it red}) 1,2 \\
CD--38:245          & $-$4.11 & Fig.~\ref{fig:NEMP}f: ({\it red}) 3,4,5,6,7,8,9,
 ({\it blue}) 17, ({\it green}) 1 \\  \hline
\multicolumn{3}{c}{UMP stars} \\ \hline
HE~0557--4840       & $-$4.77 & Fig.~\ref{fig:UMP}a: ({\it red}) 31,32 ($T_{\rm eff}=4900$K),
 ({\it blue}) 31 ($T_{\rm eff}=5100$K) ({\it green}) 1,2 \\
SDSS~J102915+172927 & $-$4.92 & Fig.~\ref{fig:UMP}b: ({\it red}) 33,34 \\ \hline
\multicolumn{3}{c}{HMP stars} \\ \hline
HE~0107--5240       & $-$5.61 & Fig.~\ref{fig:HMP}a: ({\it red}) 35,36,
 ({\it blue}) 37, ({\it green}) 1,2 \\ 
HE~1327--2326       & $-$5.88 & Fig.~\ref{fig:HMP}b: ({\it red}) 38,39,
 ({\it blue}) 38, ({\it green}) 1,2 \\ \hline
\multicolumn{3}{c}{EMP stars in dwarf galaxies} \\ \hline
Scl~031--11         & $-$3.47 & Fig.~\ref{fig:dSph}a: ({\it red}) 40 \\
Scl~07--49          & $-$3.48 & Fig.~\ref{fig:dSph}b: ({\it red}) 41 \\
SEGUE1-7            & $-$3.57 & Fig.~\ref{fig:dSph}c: ({\it red}) 42 \\
Fnx~05--42          & $-$3.66 & Fig.~\ref{fig:dSph}d: ({\it red}) 41 \\
Boo--1137           & $-$3.71 & Fig.~\ref{fig:dSph}e: ({\it red}) 43 \\
S~1020549           & $-$3.81 & Fig.~\ref{fig:dSph}f: ({\it red}) 44 \\
Scl~07--50          & $-$3.96 & Fig.~\ref{fig:dSph}g: ({\it red}) 41 \\ \hline
\multicolumn{3}{c}{MP stars in Hercules dwarf spheroidal galaxy} \\ \hline
Her-2               & $-$2.02 & Fig.~\ref{fig:Her}: ({\it red}) 45 \\
Her-3               & $-$2.04 & Fig.~\ref{fig:Her}: ({\it blue}) 45 
\enddata

\tablecomments{The columns show the name of star, the metallicity [Fe/H],
 and the figure number with references and color legends.}
\tablerefs{(1) \cite{yon13}, (2) \cite{nor13p4}, (3) \cite{cay04}, (4)
 \cite{spi05}, (5) \cite{spi06}, (6) \cite{and07}, (7) \cite{and08}, 
 (8) \cite{and10}, (9) \cite{spi12}, (10) \cite{caf11}, (11) \cite{coh04}, 
 (12) \cite{aok09}, (13) \cite{lai08}, (14) \cite{coh08}, (15)
 \cite{bon09}, (16) \cite{gon08}, (17) \cite{nor01}, (18) \cite{ish10},
 (19) \cite{ste02emp}, (20) \cite{coh06}, (21) \cite{aok04}, (22)
 \cite{fre07}, (23) \cite{bar05}, (24) \cite{ito09}, (25) \cite{ito13},
 (26) \cite{ple05}, (27) \cite{bee07}, (28) \cite{hon04}, (29)
 \cite{spi11} (30) \cite{nor02}, (31) \cite{nor07}, (32) \cite{nor12},
 (33) \cite{caf11nat}, (34) \cite{caf12}, (35) \cite{chr04}, (36)
 \cite{bes05}, (37) \cite{col06}, (38) \cite{fre08}, (39) \cite{bon12},
 (40) \cite{sta12}, (41) \cite{taf10}, (42) \cite{nor10}, (43)
 \cite{nor10boo}, (44) \cite{fre10}, (45) \cite{koc08}}
\end{deluxetable}

\clearpage

\LongTables
\begin{landscape}
\begin{deluxetable}{cccccccccccc}
 \tabletypesize{\scriptsize}
 \tablecaption{Pop III SN models with $25\Msun$ progenitor
 \label{tab:models}}
 \tablewidth{0pt}
 \tablehead{
   \colhead{Star name}
 & \colhead{$E_{51}$}
 & \colhead{$\Mfe/\Msun$}
 & \colhead{$\Mrem/\Msun$}
 & \colhead{$\MCini/\Msun$}
 & \colhead{$\Mmout/\Msun$}
 & \colhead{$\fej$}
 & \colhead{$\frho$}
 & \colhead{$\ye$ modification}
 & \colhead{Progenitor}
 & \colhead{Figure (line type)}
 & \colhead{Distribution}
 }
\startdata
\multicolumn{12}{c}{EMP stars} \\ \hline
HE~0146--1548       &  5 & $2.38\times10^{-2}$ & 3.45 & 1.78 & 3.75 &
 0.15   & 1/8  & Mn \& Co & A & Fig.~\ref{fig:EMP1}a (solid) & Yes \\
CS~22189--009       & 20 & $7.76\times10^{-2}$ & 4.20 & 1.72 & 4.51 &
 0.11   & 1/2  & Mn \& Co & B & Fig.~\ref{fig:EMP1}b (solid) & Yes \\
SDSS~J090733+024608 &  5 & $1.85\times10^{-1}$ & 1.74 & 1.74 & ---  &
 1.0    & 1/8  & No  & A & Fig.~\ref{fig:EMP1}c (solid) &  \\
CS~22897--008       & 20 & $5.65\times10^{-2}$ & 4.22 & 1.72 & 4.43 &
 0.08   & 1/2  & Mn \& Co & B & Fig.~\ref{fig:EMP1}d (solid) & Yes \\
---                 &  5 & $8.46\times10^{-2}$ & 2.58 & 1.61 & 2.99 &
 0.3    & 1/6  & Mn \& Co & B & Fig.~\ref{fig:EMP1}d (dashed) & \\
BS~16545--089       & 20 & $1.07\times10^{-1}$ & 4.02 & 1.74 & 4.59 &
 0.2    & 1/3  & Mn \& Co & A & Fig.~\ref{fig:EMP1}e (solid) &  \\
CS~22963--004       &  5 & $2.85\times10^{-2}$ & 4.75 & 1.77 & 5.28 &
 0.15   & 1/6  & Mn \& Co & B & Fig.~\ref{fig:EMP1}f (solid) & Yes \\
CS~22968--014       & 20 & $8.61\times10^{-2}$ & 3.84 & 1.70 & 4.14 &
 0.12   & 1/2  & Mn \& Co & B & Fig.~\ref{fig:EMP1}g (solid) & Yes \\
HE~1356--0622       &  5 & $1.11\times10^{-1}$ & 1.96 & 1.78 & 2.31 &
 0.65   & 1/8  & No  & A & Fig.~\ref{fig:EMP1}h (solid) &  \\
HE~1347--1025       &  5 & $9.85\times10^{-2}$ & 4.00 & 1.72 & 6.27 &
 0.5    & 1/8  & Co  & A & Fig.~\ref{fig:EMP1}i (solid) & Yes \\
BS~16076--006       &  5 & $5.10\times10^{-2}$ & 3.87 & 1.70 & 4.59 &
 0.25   & 1/8  & Co  & B & Fig.~\ref{fig:EMP1}j (solid) & Yes \\
CS~22876--032AB     & 20 & $4.24\times10^{-2}$ & 4.49 & 1.72 & 4.67 &
 0.06   & 1/2  & Mn \& Co & A & Fig.~\ref{fig:EMP2}a (solid) & Yes \\
HE~0049--3948       & 20 & $8.34\times10^{-2}$ & 4.28 & 1.66 & 4.74 &
 0.15   & 1/3  & No  & A & Fig.~\ref{fig:EMP2}b (solid) &  \\
HE~1320--2952       & 10 & $1.30\times10^{-1}$ & 2.73 & 2.00 & 3.22 &
 0.4    & 1/4  & Co  & A & Fig.~\ref{fig:EMP2}c (solid) &  \\
BS~16467--062       & 20 & $4.36\times10^{-2}$ & 4.51 & 2.25 & 4.81 &
 0.12   & 1/3  & Mn \& Co & A & Fig.~\ref{fig:EMP2}d (solid) & Yes \\
---                 &  5 & $7.44\times10^{-2}$ & 3.79 & 1.86 & 5.72 &
 0.5    & 1/6  & Mn \& Co & A & Fig.~\ref{fig:EMP2}d (dashed) & \\
CS~22172--002       & 20 & $7.76\times10^{-2}$ & 3.87 & 1.72 & 4.14 &
 0.11   & 1/2  & Mn \& Co & B & Fig.~\ref{fig:EMP2}e (solid) & Yes \\
HE~0228--4047       & 20 & $1.36\times10^{-1}$ & 3.27 & 1.63 & 3.98 &
 0.3    & 1/4  & No  & A & Fig.~\ref{fig:EMP2}f (solid) &  \\
G~238-30            &  5 & $6.80\times10^{-2}$ & 2.55 & 1.78 & 3.07 &
 0.4    & 1/8  & No  & A & Fig.~\ref{fig:EMP2}g (solid) &  \\
CS~22885--096       & 20 & $4.95\times10^{-2}$ & 3.76 & 2.12 & 3.98 &
 0.12   & 1/3  & Co  & B & Fig.~\ref{fig:EMP2}h (solid) & Yes \\ \hline
\multicolumn{12}{c}{CEMP stars} \\ \hline
HE~1150--0428       &  5 & $3.43\times10^{-4}$ & 5.83 & 1.67 & 5.84 &
 0.003  & 1/16 & No  & B & Fig.~\ref{fig:CEMP}a (solid) &  \\
---                 &  5 & $3.51\times10^{-4}$ & 5.83 & 1.67 & 5.84 &
 0.003  & 1/16 & No  & A & Fig.~\ref{fig:CEMP}a (dashed) &  \\
HE~1012--1540       & 20 & $1.11\times10^{-3}$ & 4.81 & 2.25 & 4.81 &
 0.003  & 1/3  & Co  & A & Fig.~\ref{fig:CEMP}b (solid) & Yes \\
CS~29498--043       & 20 & $9.08\times10^{-4}$ & 5.11 & 1.90 & 5.12 &
 0.0015 & 1/2  & No  & B & Fig.~\ref{fig:CEMP}c (solid) & Yes \\
---                 & 20 & $9.10\times10^{-4}$ & 5.27 & 1.90 & 5.28 &
 0.0015 & 1/2  & No  & A & Fig.~\ref{fig:CEMP}c (dashed) & \\
HE~1506--0113       & 20 & $6.80\times10^{-3}$ & 5.15 & 1.86 & 5.20 &
 0.015  & 1/4  & Mn \& Co & A & Fig.~\ref{fig:CEMP}d (solid) & Yes \\
CS~29527-015        & 20 & $6.52\times10^{-3}$ & 5.14 & 1.76 & 5.20 &
 0.015  & 1/4  & Mn \& Co & A & Fig.~\ref{fig:CEMP}e (solid) & Yes \\
HE~0132--2429       & 20 & $1.24\times10^{-2}$ & 5.39 & 1.81 & 5.50 &
 0.03   & 1/4  & Mn \& Co & B & Fig.~\ref{fig:CEMP}f (solid) & Yes \\
HE~1300+0157        & 20 & $5.21\times10^{-2}$ & 3.25 & 1.76 & 3.46 &
 0.12   & 1/4  & Mn \& Co & A & Fig.~\ref{fig:CEMP}g (solid) & Yes \\
---                 & 20 & $3.42\times10^{-1}$ & 2.16 & 2.16 & ---  &
 1.0    & 1/4  & Mn \& Co & A & Fig.~\ref{fig:CEMP}g (dashed) & \\
---                 & 20 & $3.58\times10^{-3}$ & 5.57 & 2.08 & 5.61 &
 0.01   & 1/4  & Mn \& Co & A & Fig.~\ref{fig:CEMP}g (dotted) & \\
BD+44$^\circ$493    &  5 & $6.17\times10^{-3}$ & 5.33 & 1.90 & 5.50 &
 0.045  & 1/8  & Co  & A & Fig.~\ref{fig:CEMP}h (solid) & Yes \\
HE~1201--1512       &  5 & $9.17\times10^{-3}$ & 5.65 & 1.55 & 5.80 &
 0.035  & 1/8  & Mn \& Co & A & Fig.~\ref{fig:CEMP}i (solid) & Yes \\
---                 &  5 & $4.53\times10^{-3}$ & 5.72 & 1.78 & 5.80 &
 0.02   & 1/4  & Mn \& Co & A & Fig.~\ref{fig:CEMP}i (solid) &  \\
HE~2139--5432       &  5 & $6.78\times10^{-4}$ & 5.03 & 1.62 & 5.04 &
 0.003  & 1/8  & Mn \& Co & B & Fig.~\ref{fig:CEMP}j (solid) & Yes \\
53327-2044-515      &  5 & $8.51\times10^{-3}$ & 5.56 & 1.76 & 5.76 &
 0.05   & 1/8  & Mn \& Co & A & Fig.~\ref{fig:CEMP}k (solid) & Yes \\
---                 &  5 & $3.26\times10^{-3}$ & 5.66 & 1.74 & 5.72 &
 0.015  & 1/6  & Mn \& Co & A & Fig.~\ref{fig:CEMP}k (dashed) &  \\
G~77--61            &  1 & $1.85\times10^{-4}$ & 6.17 & 1.90 & 6.18 &
 0.002  & No   & Mn \& Co & B & Fig.~\ref{fig:CEMP}l (solid) &  \\ \hline
\multicolumn{12}{c}{NEMP stars} \\ \hline
CS~22952--015       &  5 & $3.09\times10^{-2}$ & 4.46 & 1.84 & 5.12 &
 0.2    & 1/6  & Mn  & A & Fig.~\ref{fig:NEMP}a (solid) & Yes \\
BS~16550--087       & 10 & $7.68\times10^{-2}$ & 3.14 & 1.79 & 3.46 &
 0.19   & 1/4  & Mn  & A & Fig.~\ref{fig:NEMP}b (solid) & Yes \\
CS~22949--037       & 10 & $2.42\times10^{-3}$ & 4.34 & 1.66 & 4.36 &
 0.005  & 1/4  & Mn \& Co & A & Fig.~\ref{fig:NEMP}c (solid) & Yes \\
---                 & 10 & $4.85\times10^{-3}$ & 4.11 & 1.66 & 4.13 &
 0.01   & 1/4  & Mn \& Co & A & Fig.~\ref{fig:NEMP}c (dashed) &  \\
CS~30336-049        &  5 & $2.54\times10^{-2}$ & 5.01 & 1.61 & 5.35 &
 0.09   & 1/6  & Mn \& Co & A & Fig.~\ref{fig:NEMP}d (solid) & Yes \\
---                 &  5 & $1.69\times10^{-1}$ & 2.16 & 1.61 & 2.99 &
 0.6    & 1/6  & Mn \& Co & A & Fig.~\ref{fig:NEMP}d (dashed) &  \\
HE~0057--5959       &  5 & $6.80\times10^{-3}$ & 5.38 & 1.78 & 5.53 &
 0.04   & 1/8  & No  & A & Fig.~\ref{fig:NEMP}e (solid) & \\
CD--38:245          & 20 & $6.10\times10^{-2}$ & 4.07 & 1.64 & 4.28 &
 0.08   & 1/2  & Mn \& Co & A & Fig.~\ref{fig:NEMP}f (solid) & Yes \\ \hline
\multicolumn{12}{c}{UMP stars} \\ \hline
HE~0557--4840       &  5 & $6.80\times10^{-3}$ & 5.57 & 1.78 & 5.72 &
 0.04   & 1/8  & No  & A & Fig.~\ref{fig:UMP}a (solid) & Yes \\
SDSS~J102915+172927 & 20 & $1.21\times10^{-1}$ & 3.03 & 1.58 & 3.29 &
 0.15   & 1/2  & Mn \& Co & A & Fig.~\ref{fig:UMP}b (solid) &  \\
---                 &  5 & $9.25\times10^{-2}$ & 3.49 & 1.78 & 5.20 &
 0.5    & 1/6  & Mn \& Co & A & Fig.~\ref{fig:UMP}b (dashed) &  \\ \hline
\multicolumn{12}{c}{HMP stars} \\ \hline
HE~0107--5240       &  5 & $8.02\times10^{-5}$ & 6.24 & 1.90 & 6.24 &
 0.0005 & 1/4  & Mn \& Co & A & Fig.~\ref{fig:HMP}a (solid) & Yes \\
HE~1327--2326       &  5 & $1.45\times10^{-5}$ & 5.72 & 1.74 & 5.72 &
 0.00008& 1/8  & No  & B & Fig.~\ref{fig:HMP}b (solid) & Yes \\
---                 &0.72& $1.53\times10^{-5}$ & 5.72 & 1.69 & 5.72 &
 0.00006& 1/8  & No  & B & Fig.~\ref{fig:HMP}b (dashed) &  \\ \hline
\multicolumn{12}{c}{EMP stars in dwarf galaxies} \\ \hline
Scl~031--11         & 20 & $2.82\times10^{-1}$ & 3.26 & 1.72 & 4.28 &
 0.4    & 1/2  & Mn \& Co & A & Fig.~\ref{fig:dSph}a (solid) &  \\
---                 & 10 & $1.34\times10^{-2}$ & 5.80 & 1.80 & 5.92 &
 0.03   & 1/3  & Mn \& Co & A & Fig.~\ref{fig:dSph}a (dashed) &  \\
Scl~07--49          & 20 & $1.77\times10^{-1}$ & 2.63 & 1.74 & 3.22 &
 0.4    & 1/4  & No  & A & Fig.~\ref{fig:dSph}b (solid) &  \\
SEGUE1-7            & 20 & $7.55\times10^{-4}$ & 5.72 & 2.53 & 5.72 &
 0.003  & 1/4  & Co  & A & Fig.~\ref{fig:dSph}c (solid) & Yes \\
---                 & 20 & $6.99\times10^{-4}$ & 5.72 & 1.96 & 5.72 &
 0.0018 & 1/4  & Co  & A & Fig.~\ref{fig:dSph}c (dashed) &  \\
Fnx~05--42          & 20 & $1.18\times10^{-1}$ & 3.36 & 2.20 & 3.98 &
 0.35   & 1/4  & No  & A & Fig.~\ref{fig:dSph}d (solid) & Yes \\
Boo--1137           &  5 & $4.87\times10^{-2}$ & 4.65 & 1.90 & 5.72 &
 0.28   & 1/4  & Co  & A & Fig.~\ref{fig:dSph}e (solid) & Yes \\
---                 & 20 & $8.41\times10^{-2}$ & 3.30 & 1.94 & 3.60 &
 0.18   & 1/3  & Co  & A & Fig.~\ref{fig:dSph}e (dashed) &  \\
S~1020549           & 10 & $1.43\times10^{-1}$ & 2.39 & 1.94 & 2.69 &
 0.4    & 1/4  & Mn \& Co & A & Fig.~\ref{fig:dSph}f (solid) & Yes \\
Scl~07--50          & 10 & $6.96\times10^{-2}$ & 4.75 & 1.94 & 5.46 &
 0.2    & 1/4  & Co  & A & Fig.~\ref{fig:dSph}g (solid) & Yes \\ \hline
\multicolumn{12}{c}{MP stars in Hercules dwarf spheroidal galaxy} \\ \hline
Her-2               & 20 & $1.41\times10^{-2}$ & 4.97 & 1.72 & 5.04 &
 0.02   & 1/2  & Mn \& Co & A & Fig.~\ref{fig:Her} (solid) &  \\
Her-3               & 20 & $2.68\times10^{-2}$ & 4.59 & 1.74 & 4.74 &
 0.05   & 1/3  & Mn \& Co & A & Fig.~\ref{fig:Her} (dashed) & 
\enddata

\tablecomments{The numbers shown are the explosion energy,
 the ejected Fe mass, the final central remnant mass, the mass of inner
 boundary of the mixing region, the mass of outer boundary of the mixing
 region, the ejection factor, the ``low-density'' factor, the
 $\ye$ modification, the progenitor model, the figure number with line legend, and
 whether the model is used for the distribution in
 \S~\ref{sec:dist}. The masses are in unit of $\Msun$.}

\end{deluxetable}
\clearpage
\end{landscape}

\begin{deluxetable}{cccccccccc}
 \tabletypesize{\scriptsize}
 \tablecaption{Pop III SN models with $40\Msun$ progenitor
 \label{tab:40Msun}}
 \tablewidth{0pt}
 \tablehead{
   \colhead{Star name}
 & \colhead{$E_{51}$}
 & \colhead{$\Mfe/\Msun$}
 & \colhead{$\Mrem/\Msun$}
 & \colhead{$\MCini/\Msun$}
 & \colhead{$\Mmout/\Msun$}
 & \colhead{$\fej$}
 & \colhead{$\frho$}
 & \colhead{$\ye$ modification}
 & \colhead{Figure (line type)}
 }
\startdata
HE~1347--1025       & 20 & $1.44\times10^{-1}$ & 5.33 & 2.24 & 5.88 &
 0.15   & 1/5  & Mn \& Co & Fig.~\ref{fig:40Msun}a (solid) \\
CS~29527-015        & 30 & $1.88\times10^{-2}$ & 9.53 & 2.24 & 9.65 &
 0.015  & 1/4  & Mn \& Co & Fig.~\ref{fig:40Msun}b (solid) \\
BD+44$^\circ$493    & 15 & $8.39\times10^{-3}$ &12.23 & 2.47 &12.38 &
 0.015  & 1/8  & Co  & Fig.~\ref{fig:40Msun}c (solid) \\
---                 & 15 & $8.66\times10^{-3}$ &12.23 & 2.84 &12.38 &
 0.02   & 1/8  & Co  & Fig.~\ref{fig:40Msun}c (dashed) \\
HE~0557--4840       & 20 & $1.14\times10^{-2}$ &12.49 & 3.19 &12.68 &
 0.02   & 1/5  & No  & Fig.~\ref{fig:40Msun}d (solid) \\
HE~0107--5240       & 10 & $1.06\times10^{-4}$ &13.89 & 2.03 &13.89 &
 0.0003 & 1/8  & Mn \& Co & Fig.~\ref{fig:40Msun}e (solid) 
\enddata

\tablecomments{The numbers shown are the explosion energy,
 the ejected Fe mass, the final central remnant mass, the mass of inner
 boundary of the mixing region, the mass of outer boundary of the mixing
 region, the ejection factor, the ``low-density'' factor, the
 $\ye$ modification, and the figure number with
 line legend. The masses are in unit of $\Msun$.}

\end{deluxetable}

\end{document}